\PassOptionsToClass{12pt}{revtex4-1}
\documentclass[preprint]{aastex631}

\usepackage{amsmath}
\usepackage{CJKutf8}
\usepackage{stix2}
\usepackage{anyfontsize,graphicx}
\submitjournal{APJ}
\published{2024 October 8}

\newcommand{\cntext}[1]{\begin{CJK}{UTF8}{gbsn}{\rm #1}\end{CJK}\kern-1ex}

\begin{document}

\title{Flare-accelerated Electrons in the Kappa Distribution from X-Ray Spectra with
the Warm-Target Model}

\correspondingauthor{Yingjie Luo}
\email{Yingjie.Luo@glasgow.ac.uk}

\author[0000-0002-5431-545X]{Yingjie Luo (\cntext{骆英杰})} \affiliation{School
of Physics \& Astronomy, University of Glasgow, G12 8QQ, Glasgow, UK}

\author[0000-0002-8078-0902]{Eduard P. Kontar}
\affiliation{School of Physics \& Astronomy, University of Glasgow, G12 8QQ,
Glasgow, UK}

\author[0000-0002-2651-5120]{Debesh Bhattacharjee}
\affiliation{School of Physics \& Astronomy, University of Glasgow, G12 8QQ,
Glasgow, UK}

\begin{abstract}
X-ray observations provide important and valuable insights into the acceleration
and propagation of nonthermal electrons during solar flares. Improved X-ray
spectral analysis requires a deeper understanding of the dynamics of energetic
electrons. Previous studies have demonstrated that the dynamics of accelerated
electrons with a few thermal speeds are more complex than those with significantly higher speeds. To better describe
the energetic electrons after injection, a model considering energy diffusion
and thermalization effects in flare conditions (the warm-target model) has recently
been developed for spectral analysis of hard X-rays. This model has demonstrated how
the low-energy cutoff, which can hardly be constrained in cold-target
modeling, can be determined. However, the power-law form may not be the most
suitable representation of injected electrons. The kappa distribution, which is
proposed as a physical consequence of electron acceleration, has been applied successfully in RHESSI spectral analysis. In this study, we employ the
kappa-form injected electrons in the warm-target model to analyze two M-class
flares, observed by RHESSI and STIX, respectively. The best-fit results show
that the kappa-form energetic electron spectrum generates lower nonthermal
energy than the power-law form when producing a similar photon spectrum in the fit range. We also demonstrated that the fit parameters associated with the
kappa-form electron spectrum can be well determined with small
uncertainty. Further, the kappa distribution, which covers the entire electron energy range, enables the determination of key electron
properties such as total electron number density and average energy at the flare
site, providing valuable information on electron acceleration processes.

\end{abstract}
\keywords{Solar flares (1496); Solar physics (1476); Solar activity (1475);
Solar x-ray flares (1816); Active solar corona (1988)}

\section{Introduction} \label{sec:code}

Solar flares are events with an explosive release of energy and emissions that span
a wide range of electromagnetic wavelengths from radio to $\gamma$-rays. In
order to understand the acceleration mechanism and the transport of energetic
electrons, it is crucial to know where and how the energy is released
\citep{2017LRSP...14....2B}. Energetic electrons from solar flares can be
indirectly observed through X-ray and radio emissions. Because of its unique
advantages, hard X-ray (HXR) diagnostics is essential for determining the
electron properties (see \citet{2011SSRv..159..107H,2011SSRv..159..301K} for
reviews). HXR emission observed from solar flares is dominantly due to
electron-ion bremsstrahlung. The intensity of HXR emission  is linearly
proportional to the nonthermal electron density, so the observed HXR photon
spectrum provides a straightforward relationship with the energetic electrons.
Furthermore, HXR emission is less affected by propagation effects than radio
emissions, resulting in minimal modification of the original electron properties
during observations.

Over the past two decades, advanced X-ray instruments, such as the Reuven Ramaty
High-Energy Solar Spectroscopic Imager \citep[RHESSI;][]{2002SoPh..210....3L}
and the Spectrometer/ Telescope for Imaging X-rays
\citep[STIX;][]{2020A&A...642A..15K}, have facilitated the collection of
high-quality spectra and spectroscopic imaging. The frequent observation of X-ray
sources in the corona and footpoints in the chromosphere \citep{2002SoPh..210..245S,2003ApJ...595L.107E,2006A&A...456..751B,2008A&A...489L..57K,2009ApJ...691..299S}
strongly supports the 'thick-target' interpretation
\citep{1971SoPh...18..489B,1972SvA....16..273S}, where flare-accelerated
electrons propagate down to the dense chromosphere and deposit all their energy
within the chromosphere. According to the thick-target model, the observed X-ray
photon spectrum is determined by the injected electron spectrum, the bremsstrahlung
cross section, and the energy loss rate within the target
\citep{2003ApJ...595L.115B}. When the energy of electrons is high compared to that of
the target, i.e. $E \gg k_\text{B}T$ ($k_\text{B}$ is the Boltzmann constant,
$T$ is the temperature), the model is referred to as a cold target. In
cold-target conditions, the kinetic energy loss is predominantly due to Coulomb
collisions \citep{1971SoPh...18..489B,1976SoPh...50..153L}, while other
effects, such as return current ohmic losses
\citep{1977ApJ...218..306K,1980ApJ...235.1055E,2006ApJ...651..553Z,2012ApJ...745...52H,2017ApJ...851...78A},
could affect the results. While such a cold-target model is widely utilized as
the basis of HXR spectral analysis, observations suggest that the coronal
plasma during flares can reach high temperatures above ${10^7}$ K, highlighting
the need to improve the model. Studies have demonstrated that the behavior of
electrons in the warm-target condition ($E\sim k_\text{B}T$) is more complex and
significantly different from what is predicted by the cold-target model.
\citet{2003ApJ...595L.107E} highlighted a rapid drop in the energy loss rate in
warm-target conditions, indicating a substantial overestimation of electron flux
by the cold-target model in generating observed photon spectra. Additionally,
collisional energy diffusion and thermalization \citep{2005A&A...438.1107G} have
a significant impact on the dynamics of injected electrons. Furthermore, a
physically self-consistent model requires consideration of both energy diffusion and
spatial diffusion during electron transportation \citep{2014ApJ...787...86J}.
Taking into account these advances, \citet{2015ApJ...809...35K,
2019ApJ...871..225K} developed a warm-target model that provides a physically
sound treatment of electron dynamics in the warm-target condition (referred to
as the warm-target model hereafter). This model includes collisional
energy diffusion, suggesting that injected electrons thermalize within the
target rather than losing all their energy and disappearing. The HXR spectrum
generated by the warm-target model includes both the contributions from the
electrons in the cold-target chromosphere and the the component from the
corona.

The warm-target model has been utilized to analyze RHESSI flares to more
precisely fit the characteristics of nonthermal electrons
\citep{2019ApJ...871..225K,2016ApJ...832...27A,2017ApJ...836...17A}. One
significant challenge in inferring nonthermal electron properties is that the
nonthermal emission is typically masked by the thermal emissions below
20-30~keV \citep{2011SSRv..159..107H}. This makes it difficult to determine the
low-energy cutoff, $E_c$, which is important in nonthermal electron energetics
\citep[see][for a review]{2019ApJ...881....1A}.  
This parameter and the closely related nonthermal power are the key
parameters to characterize the acceleration mechanism. The cold-target condition
$E \gg k_\text{B}T$ is typically applicable in the range above 20 keV, providing
solely an upper limit for the low-energy cutoff $E_c$. On the other hand, the
warm-target model considers the thermalization of injected electrons, which
enables the electron properties to be effectively constrained, because the spectra of
thermalized electrons are proportional to the population of the injected electrons.
\citet{2019ApJ...871..225K} have demonstrated that using the warm-target model
with a power-law distribution can significantly enhance the accuracy of
determining the low-energy cutoff $E_c$ compared to the cold-target model.

However, it is important to recognize that the power-law form of the injected
electron spectrum is not unique and not always the most suitable representation
in many cases \citep{2006ApJ...643..523B}. The power-law distribution requires a
low-energy cutoff ($E_c$) to prevent an infinite number of electrons. The presence of
the low-energy cutoff in the injected electron spectrum, 
while suggested to have physical significance in terms
of the critical velocity of runaway and free acceleration
\citep{1992ApJ...400L..79H}, 
results in a nonphysical discontinuous electron distribution of a
Maxwellian plus the power law. 
Around the low-energy cutoff $E_c$, where the slope of the distribution is positive ($df/dv>0$), Langmuir waves are expected to generate and to grow \citep{1963JETP...16..682V,1964AnPhy..28..478D,1984ApJ...279..882E,2009ApJ...707L..45H}. The interaction between the waves will result in a flattened distribution around
the low-energy cutoff \citep{1973ppp..book.....K}, suggesting the injected (accelerated) 
electron distribution should be a nonincreasing function 
for all energies. On the other hand, the kappa distribution, consisting of a Maxwellian core and smoothly merged power-law tail, has been studied to represent accelerated electrons. 
The kappa distribution is often used
to model the entire energetic electron spectrum from in-situ observations
\citep{1989BAICz..40..175V,1995PhPl....2.2098M,1996GeoRL..23.1191C,2005JGRA..110.9104M,2011JGRA..116.8217I}
and has emerged as a promising alternative for describing the accelerated
electrons in HXR spectral fitting
\citep{2009A&A...497L..13K,2013ApJ...764....6O,2015ApJ...799..129O,2015ApJ...815...73B,2017ApJ...835..124E}.
Previous studies have shown that while spatially integrated spectra cannot be
well fitted only with the electrons obeying a kappa distribution, kappa-form
accelerated electrons with a prominent thermal component yield promising
results. 

In addition to not requiring an arbitrary energy cutoff $E_c$, the adoption of
a kappa-form injection offers several advantages over the power-law
distribution. First, the kappa-form electron spectrum is believed to result from
acceleration out of thermal equilibrium, and its parameters can provide crucial
information on electron acceleration. Various studies have delved into the
origin of the kappa distribution from different perspectives, including
collisional relaxation along with wave-particle interaction
\citep{1998GeoRL..25.4099M,2003ApJ...593.1134V} and beam-plasma interactions
\citep{2006JGRA..111.9106Y,2007PhPl...14j0701R,2006JGRA..111.9107R}. Moreover,
originating from the Fokker-Planck equation, \citealt{2014ApJ...796..142B} have
pointed out that the kappa electron distribution results from stochastic
acceleration in the presence of Coulomb collisions and velocity diffusion,
specifying insight into electron kinetics during solar flares. Similarly,
simulations of electron acceleration during magnetic reconnection lead to a
distribution resembling a kappa distribution \citep{2021PhRvL.126m5101A}.
Another advantage of the kappa distribution of the electron spectrum over the
power-law distribution is its entire coverage of electron kinetic energies. The kappa
distribution can provide electron information at low energy (a few keV), which
is not sensitive to X-ray instruments but can be observed through other
diagnostics such as EUV observations. This allows for a more comprehensive study
of the electron distribution through multiwavelength observations, providing
additional constraints \citep{2015ApJ...799..129O,2015ApJ...815...73B}.
Furthermore,  due to its whole electron energy coverage and convergent nature,
the kappa distribution allows for the determination of parameters such as
electron number density and average electron energy, which is not achievable
with the power-law distribution. The kappa distribution is closely related to
the thermal properties of the ambient plasma. Parameters such as electron number
density and average electron energy can be directly compared with the
corresponding thermal properties of the preaccelerated plasma, serving as an
important constraint and cross-check.

The paper investigates the spectrum of accelerated (injected) electrons in solar
flares. Section \ref{sec:nVF-kappa} discusses the relationship between the kappa-form
injected electron spectrum and the mean electron flux based on the warm-target
model. We utilize the warm-target model with kappa-form injected electrons to
analyze solar flares observed by RHESSI on 2011 February 24 and by STIX on
2022 March 28. The best-fit results and a comparison between the kappa and
power-law forms of injection is provided in Section \ref{sec:obs}. Section
\ref{sec:obs anal} discusses the uncertainty of fitting parameters related to
the kappa distribution with different approaches. The resulting characteristics
such as electron number density and average electron energy are also shown. The
findings and the physical significance of the derived kappa parameters are
summarized and discussed in Section \ref{sec:discussion}.

\section{Source-Integrated spectrum from kappa-form injected electrons}
\label{sec:nVF-kappa}

\citet{2003ApJ...595L.115B} highlighted the concept of the source-integrated
density-weighted mean electron flux $\langle nVF \rangle(E) \,[\rm{electrons\,
cm^{-2}s^{-1}keV^{-1}}]$ and expressed the observed HXR spectrum
$I(\varepsilon)\,[\rm{photons\, cm^{-2}s^{-1}keV^{-1}}]$ as:
\begin{equation}
I(\varepsilon)=\frac{1}{4\pi R^2}\int_\varepsilon^{\infty} Q(\varepsilon,E)\langle nVF \rangle(E)\, dE \label{equ: photon spec},
\end{equation}

\noindent where $R$ is the distance between the observer and the Sun, and
$Q(\varepsilon,E)$ is angle-averaged bremsstrahlung cross section (see
discussion in \citealt{2011SSRv..159..301K} for RHESSI). Determination of the mean
electron flux from the observed photon spectrum requires only the well-studied
bremsstrahlung cross section \citep{1998SoPh..178..341H}. To obtain the injected
electron rate spectrum $\dot{N}(E) \,[\rm{electrons \,s^{-1}keV^{-1}}]$ from the
obtained mean electron flux, a model for the dynamics of injected electrons is
necessary. The commonly used cold-target model establishes the relationship
between mean electron flux and injected electron rate as 
\begin{equation}
\langle nVF \rangle(E) = \frac{E}{K} \int_{E}^{\infty} \dot{N}(E_0)\,dE_0\,,\label{equ: cold}
\end{equation}
were $K = 2 \pi e^4 \ln(\Lambda)$ is the collision parameter, $e \,[\rm{esu}]$ is
the elementary charge and $ln(\Lambda)$ is the Coulomb logarithm
\citep{1962pfig.book.....S}. To account for the electron dynamics in the
warm-target corona and cold chromosphere, \citet{2015ApJ...809...35K} included
the energy diffusion, transport, and thermalization of the accelerated electrons,
rewriting Equation (\ref{equ: cold}) as
\begin{equation}\label{equ: nvf-f-ori}
\langle nVF \rangle(E) = \frac{1}{2K} E \ e^{-E/k_\text{B}T} \ \int_{E_{min}}^{E} \frac{e^{E'/k_\text{B}T}}{E'G(\sqrt{E'/k_\text{B}T})}\,dE' \times \int_{E'}^{\infty} \dot{N}(E_0)\,dE_0\,.
\end{equation}

\citet{2019ApJ...871..225K} further simplify equation (\ref{equ: nvf-f-ori}) and
rewrite it as:
\begin{equation}
\langle nVF \rangle(E) \approx \Delta EM \sqrt{\frac{8}{\pi m_e}} \frac{E}{(k_\text{B} T)^{3/2}} e^{-E/k_\text{B}T} + \frac{E}{K} \int_{E}^{\infty} \dot{N}(E_0)\,dE_0\,, \label{equ: warm}
\end{equation}
where $\Delta EM$, referred to as the 'thermalized emission measure', is due to the
thermalization of accelerated electrons and can be explicitly written as 
\begin{equation} \label{equ: add em}
\Delta EM \approx \frac{\pi}{K}\sqrt{\frac{m_e}{8}}(k_\text{B} T)^2\frac{\dot{N}_0}{E_\text{min}^{1/2}}\,,
\;\;\; \mbox{where}\;\;\;  E_\text{min} \approx 3 k_\text{B}T(\frac{5\lambda}{L})^4\,,
\end{equation}
and $\dot{N}_0 \,[\rm{electrons \,s^{-1}}]= \int_{0}^{\infty}$$\dot{N}(E) \,dE$
is the total electron injection (acceleration) rate and
$\lambda=(k_\text{B}T)^2/2Kn$ is the collisional mean free path. The half-loop
length $L$, number density $n$, and temperature $T$ are the parameters
associated with the coronal part of the loop. Equation \ref{equ: warm} consists
of two parts, referred to as the thermalized and nonthermal components, corresponding
to the warm-target (coronal) and cold-target conditions, respectively. It is worth
noting that the thermalized component is proportional to the total injected
electron rate $\dot{N}_0$, which is why this model is effective in constraining
the nonthermal component when fitting the observed X-ray spectrum. This
warm-target model \textit{'f\_thick\_warm'}, which utilizes Equations \ref{equ:
warm} and \ref{equ: add em}, is implemented in the Solar Software (SSW) and
\texttt{OSPEX} package \citep{2002SoPh..210..165S,2020ascl.soft07018T}. The function
\textit{'f\_thick\_warm'} utilizes the commonly employed broken power-law form
of injected electrons, and it is worth noting that the warm-target model
(Equations \ref{equ: warm} and \ref{equ: add em}) can accept any form of electron
injection spectrum. As mentioned earlier, the use of a kappa distribution for
injection is a promising alternative and has been applied to fit the RHESSI
spectra.

\citealt{2014ApJ...796..142B} derive the kappa distribution as a result of
stochastic acceleration in collisional plasma:
\begin{equation}\label{equ: kappa-dis-v2}
f_k(v)=\frac{n_k}{\pi^{3/2}v_\text{te}^3\kappa^{3/2}}\frac{\Gamma(\kappa)}{\Gamma(\kappa-3/2)}(1+\frac{v^2}{\kappa v_\text{te}^2})^{-\kappa}\,,
\end{equation}
where $v_\text{te}$ is the thermal speed ($1/2 m_e v_\text{te}^2=k_\text{B}
T_{\kappa}$), and $\kappa$ is the kappa index, which importantly represents the
ratio between the timescales of acceleration $\tau_\text{acc}(v)$ and collisional deceleration
$\tau_{c}(v)$, $\kappa=\frac{\tau_\text{acc}(v)}{2\tau_{c}(v)}$.
While different formats of the kappa distribution exist \citep[see,
e.g.][]{2009A&A...497L..13K,2013ApJ...764....6O}, they are identical in essence
and differ only in their format and use of parameters. Furthermore, the physical
interpretation of $\kappa$ varies when additional factors are taken into
account. In this study, we have chosen to adopt Equation \ref{equ: kappa-dis-v2}
for the kappa distribution.
\begin{figure*}[!ht]
\centering
\includegraphics[width=0.9\textwidth]{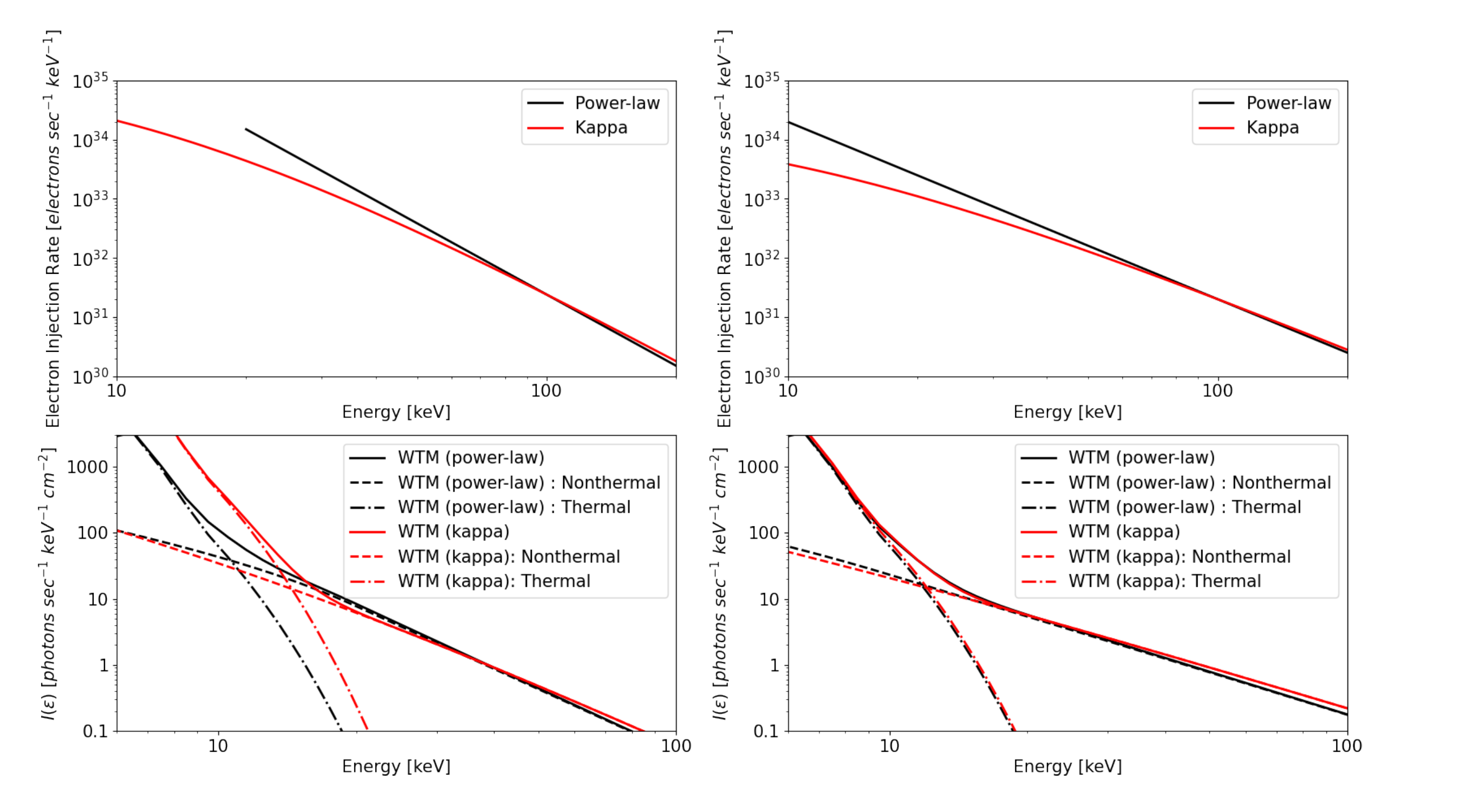}
\caption{Top Panel: the electron injection rate spectrum from distributions of the power-law form
and kappa form. For the case shown on the left, the power-law form of injection has
parameters of $\dot{N}_0=1\times10^{10} \,\rm{electrons \,s^{-1}}$, $E_c=20
\,\rm{keV}$, and $\delta=4$, and the kappa form of injection is set to
$\dot{N}_0=6.8\times10^{10} \,\rm{electrons \,s^{-1}}$, $k_\text{B} T_\kappa=1.5
\,\rm{keV}$, and $\kappa=\delta+1=5$. The selection is for satisfying
$\dot{N}_{\kappa}(E_1)=\dot{N}_{Max}(E_1)$, when $E_1=100 \,\rm{keV}$. The case
shown in the right panel is with the power-law parameters: $\dot{N}_0=1\times10^{10} \,\rm{electrons \,s^{-1}}$, $E_c=10 \,\rm{keV}$,
$\delta=3$, condition of a high proportion of
nonthermal electrons. The same prescription is used to obtain the kappa-form electron spectrum: $\dot{N}_0=1.17\times10^{10}
\,\rm{electrons \,s^{-1}}$, $k_\text{B} T_\kappa=1.5 \,\rm{keV}$, and $\kappa=4$. Bottom Panels: the photon spectrum, based on the
warm-target model (WTM), generated by the electron injection in the top panel
with a coronal condition of $L_\text{loop} = 30\,\rm{Mm}$, $n_\text{loop} =
5\times 10^{10} \,\rm{cm}^{-3}$, and $k_\text{B} T_\text{loop}=1.5 \,\rm{keV}$.
Photon spectra from power-law and kappa form of injection are plotted in black and
red, respectively. Solid curves represent the total photon spectrum, while the
dashed and dotted-dashed curves represent the 'nonthermal' and 'thermalized'
components, respectively. \label{fig: kap}}
\end{figure*}

The relationship between the injected electron rate spectrum and the electron
distribution for velocity is $\dot{N}(E)dE=A v f_k(v) d^3v$, where $A$ is the
injection area. For an isotropic electron distribution ($d^3v=4\pi v^2dv$), the
injected electron rate spectrum becomes (see red curves in Figure \ref{fig:
kap}):
\begin{equation}\label{equ: kappa-flux}
\dot{N}(E)=A\frac{n_k\Gamma(\kappa)}{\Gamma(\kappa-3/2)\kappa^{3/2}}\sqrt{\frac{8}{\pi m_ek_\text{B}T_{\kappa}}}\frac{E/k_\text{B}T_{\kappa}}{(1+E/\kappa k_\text{B}T_{\kappa})^\kappa}\,, 
\end{equation}
where $n_k\ [\rm{cm^{-3}}]= \int$$f_k(v)\,d^3v$ is the total electron number
density and $\Gamma(x)$ is the gamma function. The total injection rate becomes
\begin{equation}
\dot{N}_0= \int_{0}^{\infty} \dot{N}(E) \,dE = 2A n_k \sqrt{\frac{2k_\text{B}T_{\kappa}}{m_e}}\frac{\kappa^{1/2}}{(\kappa-2)B(\kappa-3/2,1/2)},\label{equ: kappa-ele-spec}
\end{equation}
where $B(x,y)$ is the beta function.  
Equation \ref{equ: kappa-flux} can be expressed as
\begin{equation} \label{equ: kappa-flux-2}
\dot{N}(E)=\frac{\dot{N}_0}{k_\text{B}T_{\kappa}}\frac{(\kappa-1)(\kappa-2)}{\kappa^2}\frac{E/k_\text{B}T_{\kappa}}{(1+E/\kappa k_\text{B}T_{\kappa})^\kappa}. 
\end{equation}

According to Equation \ref{equ: warm}, the mean electron flux from a kappa-form
injected electrons is given by \citep[Equation 17 in the Appendix
of][]{2019ApJ...871..225K}:

\begin{equation}
\langle nVF \rangle(E) \approx \Delta EM \sqrt{\frac{8}{\pi m_e}} \frac{E}{(k_\text{B} T)^{3/2}} e^{-E/k_\text{B}T} + \dot{N}_0\frac{E}{K} \frac{1+(1-1/\kappa)E/k_\text{B}T_\kappa}{(1+E/\kappa k_\text{B} T\kappa)^{\kappa-1}}. \label{equ: warm-kappa}
\end{equation}

For high-energy electrons, the kappa  distribution can be well approximated by a
power-law distribution with a spectral index $\kappa-1$. Figure \ref{fig: kap}
shows that the two distributions have an identical injection rate near 100~keV.
However, in the energetically important deka-keV range, the kappa distribution is
distinct from a power-law distribution, and thus a different photon spectrum
emerges. Figure \ref{fig: kap} illustrates the electron injection rate (top
panels, red and black for kappa and power-law injection, respectively) and the
resulting photon spectra (bottom panels) generated by the warm-target model in a
typical flare coronal condition ($L_\text{loop} = 30 \,\rm{Mm}$, $n_\text{loop}
=  5\times 10^{10} \, \rm{cm}^{-3}$, and $k_\text{B} T_\text{loop}=1.5
\,\rm{keV}$). 

\section{Application of the Warm-target Model to Observations} \label{sec:obs}

In this section, we apply the warm-target model to analyze two well-observed
flare events: one on 2011 February 24 observed by RHESSI, and one on 2022
March 28 observed by STIX. Both events are situated in the limb and
provide spectroscopic images with two footpoint and loop-top sources. The
warm-target model with kappa-form electron injection (function
\textit{f\_thick\_warm\_kappa}, based on Equation \ref{equ: warm-kappa})
encompasses parameters of the kappa distribution (total electron injection rate
$\dot{N}_0$, kappa temperature $T_\kappa$, and kappa index $\kappa$) and thermal
properties of the target coronal loop (the half-loop length $L$, number density
$n_\text{loop}$, and temperature $T_\text{loop}$). 

In this study, the thermal properties of the target plasma are determined by the
HXR observations of the thermal component. Following the approach outlined in
\citet{2019ApJ...871..225K}, the thermal component of the X-ray spectra prior to
the HXR burst can provide the plasma temperature ($T$) and emission measure
($EM$). Moreover, the HXR spectroscopic imaging can estimate the coronal loop
length ($L$) and the size of the coronal X-ray loop source for determining the electron
number density ($n_\text{th}$). While this method is straightforward, we recognize
that other methods, such as differential emission measure analysis \citep[DEM;
see algorithms
from][]{1976A&A....49..239C,2012A&A...539A.146H,2015ApJ...807..143C} from the EUV
flare context or combining X-ray data with EUV data
\citep{2014ApJ...789..116I,2015ApJ...815...73B}, can yield thermal properties
\citep{2013ApJ...779..107B,2015ApJ...815...73B}. 

\subsection{2011 February 24 RHESSI flare} \label{sec:20110224}

\begin{figure*}[!ht]
\plotone{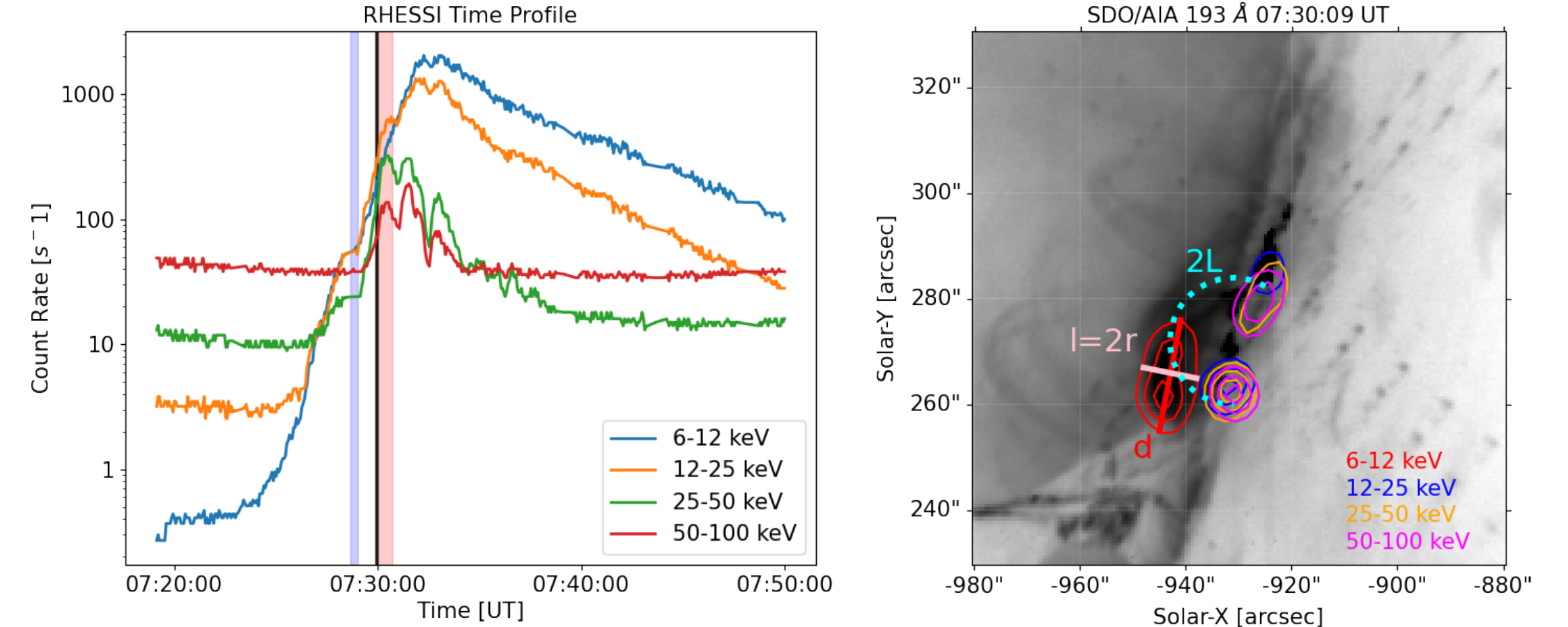}
\caption{Left panel: RHESSI light curves detailing a GOES-class M3.5 solar flare
on 2011 February 24. The blue, black, and red shaded regions denote the
chosen time frames for background, preburst, and HXR burst, respectively, for
the subsequent spectral fit. Right panel: RHESSI contours (50, 70, 90\% of the
maximum; the time range used for imaging is the red shaded region in the left
panel) overlaid on the SDO/AIA 193 \AA\ flare context (inverted grayscale). A cyan-colored semicircular loop (half-loop length $L$) that traverses
through two footpoints and the loop-top source is plotted. The 50\% contour of
the loop-top source also provides the characteristic lengths ($l=2r$ and $d$)
necessary for calculating the size of the loop-top source. } \label{fig: rhe-img}
\end{figure*}

\begin{figure*}[!ht]
\plotone{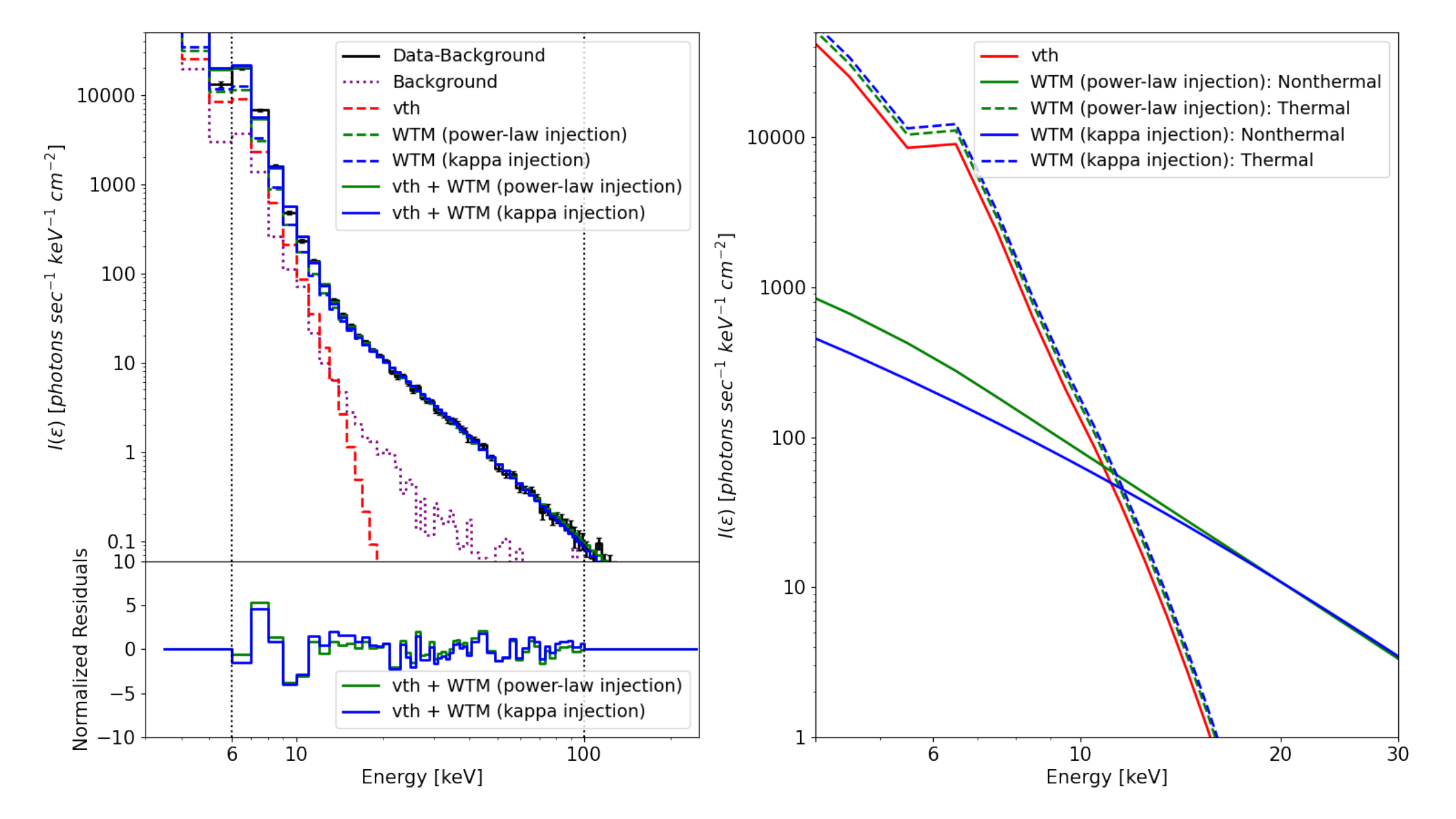}
\caption{Left panel: fit photon spectrum of the 2011 February 24 flare. The fit was done twice, using functions \textit{f$\_$vth}
(isothermal, red dashed curve) and
\textit{f$\_$thick$\_$warm}/\textit{f$\_$thick$\_$warm$\_$kappa} (warm-target model with
power-law and kappa forms of electron injection, green and blue dashed curves,
respectively). The observed background-subtracted spectrum is shown in black
(times of burst and background are shown as red and blue shaded regions in
Figure \ref{fig: rhe-img} left panel), and the fit range spans from 6 to 100
keV. Normalized residuals, which are the difference between observed (black
curve) and fit (green and blue solid curves for different fits) photons divided
by error, are also displayed in the bottom. Right panel: the photon spectra of
the isothermal (red) and warm-target models (power-law and kappa forms of injection
are in green and blue, respectively). The nonthermal and thermal components (see
Equation \ref{equ: warm}) of the warm-target model are presented as solid and
dashed curves, respectively. \label{fig: 20110224_fit}}
\end{figure*}

\begin{figure*}[!ht]
\plotone{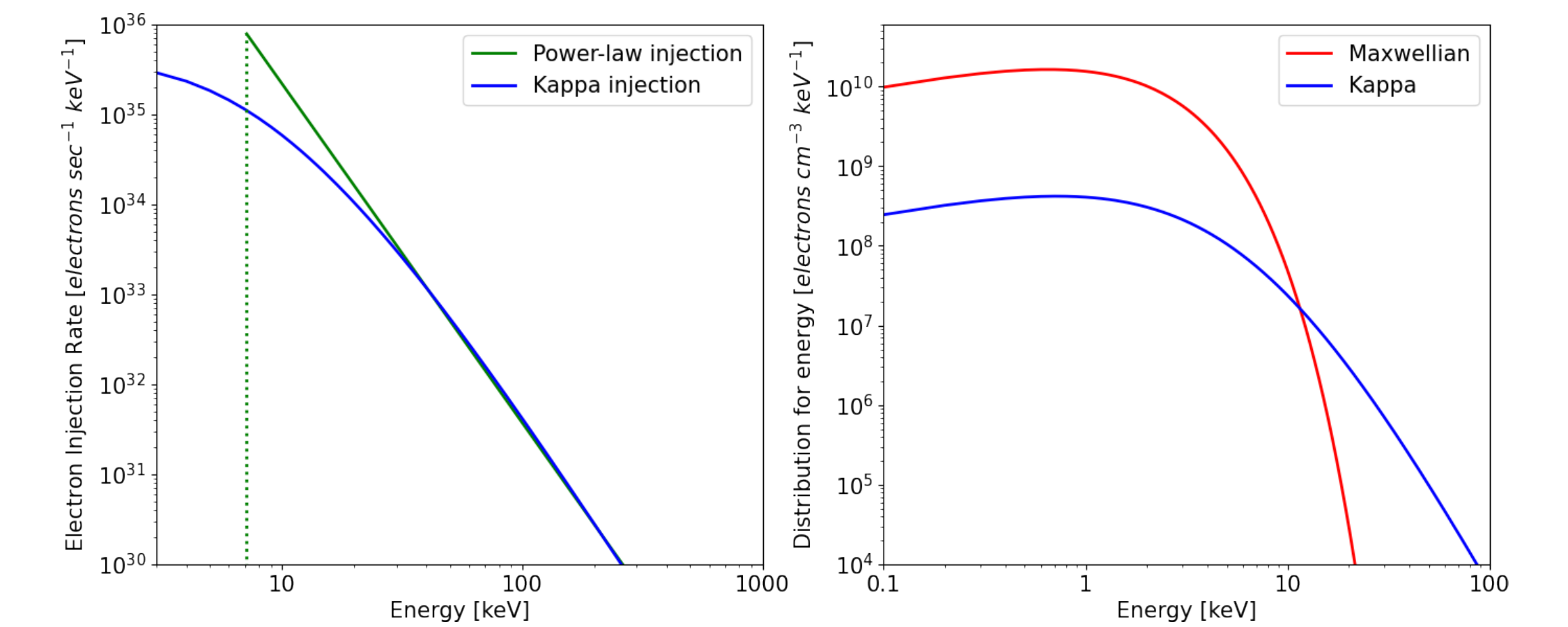}
\caption{Left panel: the injected electron rate spectrum $\dot{N}(E)$ inferred
from the warm-target fit results (see Table \ref{tab: 20110224 fit} for fitted
parameters). The green and blue curves correspond to the power-law ($E_c=7.13
\,\rm{keV}$) and kappa forms of electron injection, respectively. Right panel: the
distribution of electrons for the energy $f_k(E)$ ($f_k(E)dE = f_k(v) d^3v$) of
the fitted kappa distribution (blue curve, total electron number density
$n_k=1.67\times10^{9} \,\rm{cm}^{-3}$, from Equation \ref{equ: kappa-ele-spec})
and preburst thermal Maxwellian distribution (red curve, total electron number
density $n_k=n_{th}=4.4\times10^{10} \, \rm{cm}^{-3}$). \label{fig:
20110224_inj}}
\end{figure*}

\begin{table}
    \centering
    \begin{tabular}{||c|c|c|c||}
        Parameters & vth+WTM(power-law)  & vth+WTM(kappa) &\\
        \hline
        EM [$10^{49} \,\rm{cm}^{-3}$] & 0.127 & 0.127 & Fixed\\
        $k_\text{B} T$ [keV] & 1.30 & 1.30 & Fixed\\
        $n_\text{loop}$ [$10^{10} \,\rm{cm}^{-3}$] & 4.4 & 4.4 & Fixed\\
        $k_\text{B} T_\text{loop}$ [keV] & 1.30 & 1.30 & Fixed\\
         $L$ [Mm] & 15.8 & 15.8 & Fixed\\
        $\dot{N}_0$ [$10^{35} \,\rm{electrons} \,\rm{sec}^{-1}$] & 20.3 $\pm$
        0.8 & 22.5 $\pm$ 0.8 & Free\\
        $\delta$ & 3.77 $\pm$ 0.02 & N/A & Free\\
        $\kappa$ & N/A & 5.14 $\pm$ 0.04 & Free\\
        $E_{c}$ [keV] & 7.13 $\pm$ 0.19 & N/A & Free\\
        $k_\text{B} T_{\kappa}$ [keV] & N/A & 1.29 $\pm$ 0.04 & Free\\
        $\chi^2$ & 1.93 & 2.11 & N/A\\
        Power [$10^{28} \,\rm{erg/s}$] & 3.62 & 2.23 for total; 1.41 above $E_c$
        & N/A\\
    \end{tabular}
    \caption{The parameters and fit results of 2011 February 24 M3.5 flare using the warm-target model. The isothermal component and thermal properties of the warm-target model are fixed, and properties associated with the accelerated electron spectrum are set free. The reduced $\chi^2$ and nonthermal electron power ($P$) are also listed. For the convenience of comparison, the nonthermal power produced by the kappa-form electron spectrum above the power-law low-energy cutoff $E_c$ is also shown.}
    \label{tab: 20110224 fit}
\end{table}

The first flare we study is a GOES-class M3.5 flare located on the eastern limb
from the Earth's perspective. The flare occurred on 2011 February 24 and the GOES
soft X-ray flux peaked at $\sim$07:35 UT \citep[see][for previous studies of
this event]{2011A&A...533L...2B,2012ApJ...760..142B,2018A&A...612A..64S}. Our
attention is drawn to the first HXR peak at approximately 07:30 UT (red shaded
time range in Figure \ref{fig: rhe-img} left panel). The HXR spectroscopic
imaging of the burst displays a 'classical' configuration of the loop-top source
and two HXR footpoints (Figure \ref{fig: rhe-img}, right panel). The X-ray
observations from RHESSI of the selected HXR burst benefit from the
low/negligible pileup effect. Furthermore, RHESSI was in optimal working
condition during the flare, providing more functional detectors.

For spectral analysis, we first gather the thermal properties of the target
coronal loops according to the images of the burst (07:30:00 to 07:30:44 UT,
Figure \ref{fig: rhe-img}, left panel red shaded time range) and spectra before
the interval analyzed (short time interval just before the burst analysed,
refer as 'preburst' hereafter, in this case 07:29:52 to 07:30:00 UT,
Figure \ref{fig: rhe-img} left panel, black shaded time range). In the analysis
of the preburst X-ray spectra, we utilize functions \textit{f$\_$vth}
(isothermal) and \textit{f$\_$thick2} (cold thick target). At this stage, we
focus solely on the thermal plasma parameters and can overlook the constraints
associated with using the cold-target model. \textit{f$\_$vth} from the best-fit
result of the preburst spectra provides us with the plasma temperature
($k_\text{B}T_0=1.30 \,\rm{keV}$) and emission measure ($EM_0=0.127\times10^{49}
\,\rm{cm}^{-3}$) before the injection/acceleration of nonthermal electrons.
Additionally, we use X-ray spectroscopic imaging (Figure \ref{fig: rhe-img}
right panel) to estimate the size of the X-ray source and target coronal loop.
We determine the source size in various dimensions (see pink and red solid lines
in Figure \ref{fig: rhe-img} right panel, $r=l/2=3.7\, \rm{Mm}$ and $d=15.3 \,
\rm{Mm}$, respectively) by considering the 50\% contour of X-ray loop-top
sources. The cross-area can be calculated as $A=\pi r^2=4.21\times10^{10}
\,\rm{cm}^{2}$, and the source volume can be obtained through the cylindrical
formula: $V=A\times d=\pi r^2 d=6.46\times10^{26} \,\rm{cm}^{3}$. Thus the
thermal electron number density can be estimated using
$n_\text{loop}=\sqrt{EM_0/V}=4.4\times10^{10} \,\rm{cm}^{-3}$. To obtain the
coronal loop length, we construct a semicircular loop (see cyan dashed line in
Figure \ref{fig: rhe-img} right panel) that connects two footpoints and passes
through the loop-top source. Consequently, we determine the thermal properties:
loop temperature ($k_\text{B} T_\text{loop}=k_\text{B}T_0=1.30 \,\rm{keV}$),
loop number density ($n_\text{loop}=\sqrt{EM_0/V}=4.4\times10^{10}
\,\rm{cm}^{-3}$), and the the half-loop length ($L=15.8 \,\rm{Mm}$).

Once the thermal properties are determined, we proceed to fit the spectra of the
burst time interval (07:30:00 to 07:30:44 UT) at 6--100 keV using functions
\textit{f$\_$vth} and \textit{f\_thick\_warm\_kappa}. In the function
\textit{f$\_$vth}, the temperature and emission measure are fixed according to
the preburst spectra ($T=T_0$, $EM=EM_0$). The warm-target algorithm proposes
that the emission measure during the burst interval is a combination of the
emission measure from the preburst and the emission measure of thermalized
injected electrons ($EM=EM_0+\Delta EM$; the photon spectra associated with $EM_0$
and $\Delta EM$ are plotted as the red solid curve and the dashed curves in Figure
\ref{fig: 20110224_fit} right panel). Since the plasma is slowly changing during
the 40 s interval (electron diffusion time along the loop $\sim$56 s,
use Equation 11 of \citealt{2019ApJ...871..225K}), the parameters in
\textit{f\_thick\_warm\_kappa} related to thermal properties of the loop are
fixed in the fit (the half-loop length $L$, number density $n_\text{loop}$,
temperature $T_\text{loop}$, and the relative elemental abundances, which are 1 by default; \citealt{2013ApJ...763...86L}). The parameters of the kappa-form injection
spectrum (total electron injection rate $\dot{N}_0$, kappa temperature
$T_\kappa$, and kappa index $\kappa$, except for the high-energy cutoff
$E_\text{high}$ fixed to $10^4 \,\rm{keV}$) are determined from the fit. Moreover,
a fit with a single power-law electron injection in the warm-target model
\textit{'f\_thick\_warm'} was performed for comparison: the total injection rate
$\dot{N}_0$, power-law index $\delta$, and low-energy cutoff $E_c$ are free,
but the break energy $E_\text{break}$ and high-energy cutoff $E_\text{high}$ are
fixed to $10^4 \,\rm{keV}$. The best-fit parameters (see Table \ref{tab:
20110224 fit}) and generated photon spectra are plotted in Figure \ref{fig:
20110224_fit} (blue and green curves for warm-target model with kappa and
power-law forms of injection, respectively, same color scheme for the
normalized residuals in the bottom panel).

The fits using the warm-target model, with kappa and power-law forms of electron
injection, both produce favorable fit results with reasonable parameters. With
the systematic uncertainty of 0.02, which is the default for RHESSI in
\texttt{OSPEX}, the reduced $\chi^2$ is 1.93 and 2.11 for the fits with kappa and
power-law forms of electron injection, respectively. From Figure \ref{fig:
20110224_fit} bottom left panel, the significant normalized residuals remain
below 12 keV, where thermal bremsstrahlung dominates. This may arise from the
use of the isothermal assumption \textit{f$\_$vth} for the thermal emission
\citep{2015A&A...584A..89J}.
From the photon spectrum made by an individual component (Figure \ref{fig:
20110224_fit} right panel), we found that for this HXR burst, the emission
measure contributed by the thermalized injected electrons ($\Delta EM$, dashed
curves) is of comparable magnitude to the isothermal component ($EM$, red solid
curve). Here, we note that although the ratio of $\Delta EM/EM$ is affected by
the preburst timing chosen, all reasonable selections of preburst timing can
yield acceptable fit results. The different ratios only arise from the
identification of preaccelerated plasma. 

In Figure \ref{fig: 20110224_inj}, left panel, we plot the injected electron rate
spectrum (blue and green curves represent the best-fit kappa and power-law forms of
injection, respectively). We noticed that the power-law and kappa forms of electron
injection exhibit similarly between 40 and 100 keV but behave differently below approximately 30 keV. At the power-law low-energy cutoff $E_c$, the
electron injection rate $\dot{N}(E_c)$ of the kappa distribution is almost one
order of magnitude smaller ($1.11$ vs. $7.88$ $\times10^{35} \,\rm{electrons}
\,\rm{sec}^{-1} \,\rm{keV}^{-1}$ for the kappa and power-law forms of injection,
respectively). Additionally, the fitted injected electron spectrum in the kappa
distribution decreases more rapidly ($\kappa>\delta+1$) at higher energy.
Although the best-fit electron spectra in kappa and power-law forms have
similar total injection rates ($\dot{N}_0$ = $22.5$ vs. $20.3$ $\times10^{35}
\,\rm{electrons} \,\rm{sec}^{-1}$ for kappa and power-law forms of injection,
respectively), the power-law electron spectrum produces a more substantial
nonthermal power than the kappa-form electron spectrum (see Table \ref{tab:
20110224 fit}, $3.62$ vs $2.23$ $\times10^{28} \,\rm{erg/s}$,
$\frac{P_\kappa}{P_{power-law}} \approx 0.61$). Moreover, for kappa form
electron spectrum, the nonthermal power produced by the electrons above the
power-law low-energy cutoff $E_c$ is $1.41$ $\times10^{28} \,\rm{erg/s}$. The
equations used for calculating the power are listed below. For power-law
distribution:

\begin{equation}
P=\int_{E_c}^{\infty} E'N(E') \,dE'=\frac{\delta-1}{\delta-2}E_C\dot{N}_0. \label{equ: power power_law}
\end{equation}

\noindent For the kappa distribution:

\begin{equation}
\begin{split}
P & = \int_{E_1}^{E_2} E'N(E') \,dE' = \frac{\dot{N}_0}{(k_\text{B}T_{\kappa}\kappa)^2(\kappa-3)}(k_\text{B}T_{\kappa}\kappa+E'')\times\\
& (\frac{E''}{k_\text{B}T_{\kappa}\kappa}+1)^{-\kappa}[2(k_\text{B}T_{\kappa}\kappa)^2+2k_\text{B}T_{\kappa}(\kappa-1)\kappa E''+(\kappa^2-3\kappa+2)E''^2]\Big|_{E_1}^{E_2}. \label{equ: power kappa}
\end{split}
\end{equation}

\noindent Thus

\begin{equation}
P=\int_{0}^{\infty} E'N(E') \,dE'= \frac{2\dot{N}_0k_\text{B}T_\kappa \kappa}{\kappa-3}; \label{equ: power kappa1}
\end{equation}

\begin{equation}
\begin{split}
P & = \int_{E_c}^{\infty} E'N(E') \,dE'= \frac{\dot{N}_0}{(k_\text{B}T_{\kappa}\kappa)^2(\kappa-3)}(k_\text{B}T_{\kappa}\kappa+E_c)\times\\
& (\frac{E_c}{k_\text{B}T_{\kappa}\kappa}+1)^{-\kappa}[2(k_\text{B}T_{\kappa}\kappa)^2+2k_\text{B}T_{\kappa}(\kappa-1) \kappa E_c+(\kappa^2-3\kappa+2)E_c^2]. \label{equ: power kappa2}
\end{split}
\end{equation}

\noindent We also include a distribution plot ($f_k(E)$, $f_k(E)dE=\langle
f_k(v) \rangle d^3v$) of the accelerated kappa-form electrons (blue curve in
Figure \ref{fig: 20110224_inj} right panel, $n_k=1.67\times10^{9}
\,\rm{cm}^{-3}$, calculated from Equation \ref{equ: kappa-ele-spec}; here $A=4.21\times10^{10}
\,\rm{cm}^{2}$). The
Maxwellian thermal distribution, which is taken from the isothermal component
\textit{f$\_$vth} (total electron number density $n_k=n_\text{th}=4.4\times10^{10} \,
\rm{cm}^{-3}$, $k_\text{B}T=1.30$~keV), is also shown as a red curve for
comparison. We will provide a more detailed discussion of the parameters
obtained in the next section.

\subsection{2022 March 28 STIX flare} \label{sec:20220328}

\begin{figure*}[!ht]
\plotone{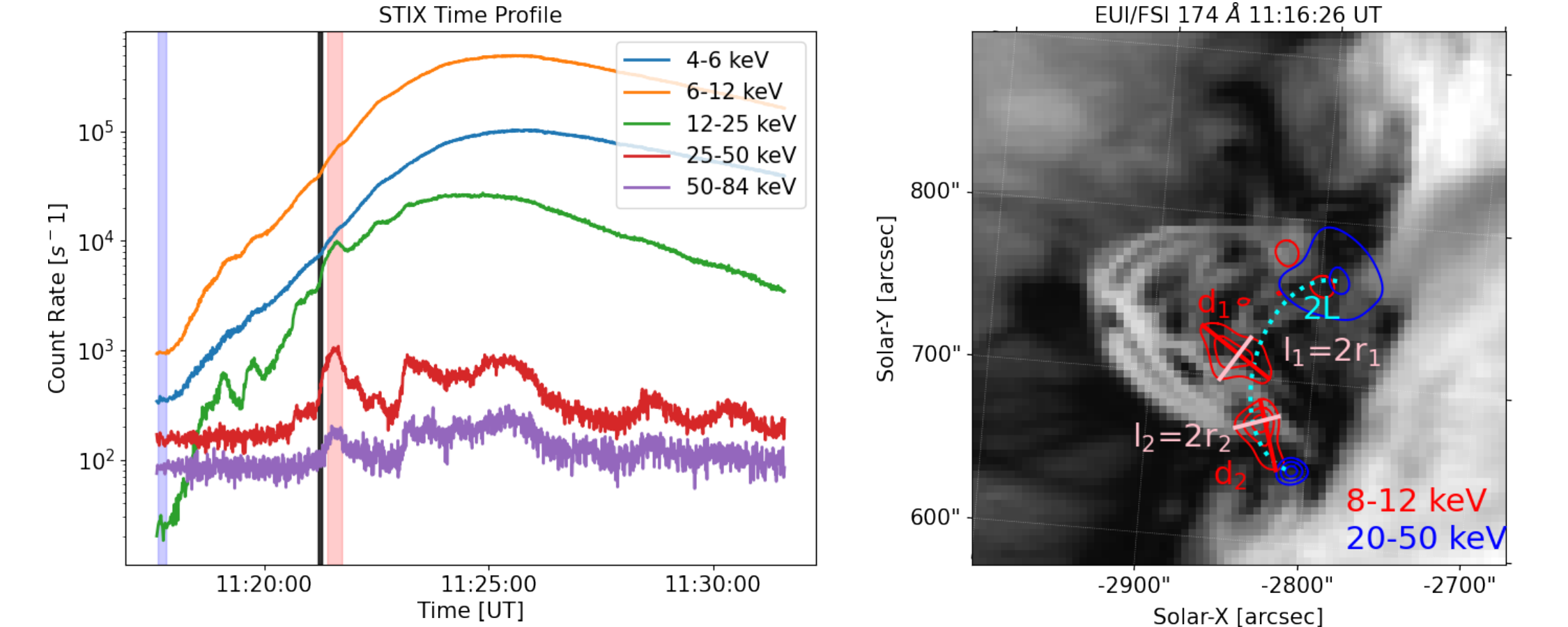}
\caption{Left panel: STIX light curves (time has been shifted to the estimated
timing observed on Earth) detailing a GOES-class M4.0 solar flare peaking at
$\sim$11:29 UT on 2022 March 28. The blue, black, and red shaded regions
denote the chosen time frames for background, preburst, and HXR burst we study,
respectively, for the subsequent spectral fit. Right panel: STIX contours (50,
70, 90\% of the maximum; the time range used for imaging is the red shaded region
in the left panel) overlaid on the EUI/FSI 174 \AA$~$flare context (inverted grayscale, using the image at 11:16 UT instead of 11:26 UT to avoid the
overexposure of the flare site). A cyan-colored semi-elliptical loop (half-loop
length $L$) that traverses through two footpoints and the loop-top source is
plotted. The 50\% contour of the two loop-top sources provides the characteristic
lengths ($l=2r$ and $d$, northern and southern loop-top sources marked with '1'
and '2', respectively) necessary for calculating the size of the loop-top source. }
\label{fig: stix-img}
\end{figure*}

\begin{figure*}[!ht]
\plotone{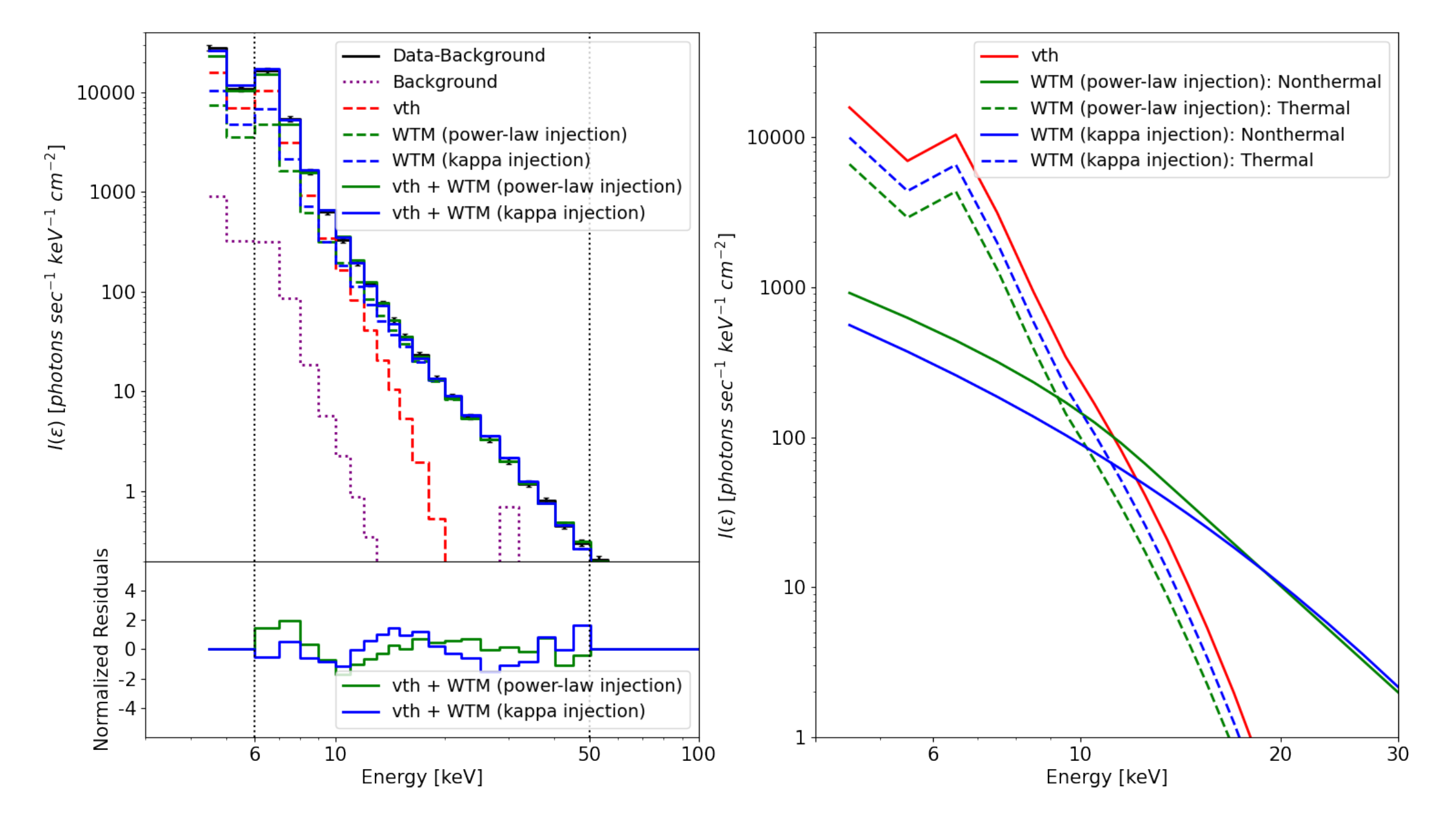}
\caption{Left panel: fit photon spectrum of the 2022 March 28 M4.0 flare. As in Figure \ref{fig: 20110224_fit}, the fit was done
twice, using functions \textit{f$\_$vth} (isothermal, red dashed curve) and
\textit{f$\_$thick$\_$warm}/\textit{f$\_$thick$\_$warm$\_$kappa} (warm-target model with
power-law and kappa forms of electron injection, green and blue dashed curves,
respectively). The observed background-subtracted spectrum is shown in black
(times of burst and background are shown as red and blue shaded regions in
Figure \ref{fig: stix-img} left panel), and the fit range spans from 6 to 50
keV. Normalized residuals are displayed at the bottom. Right panel: the photon
spectrum of the isothermal (red) and warm-target models (power-law and kappa
injection, in green and blue, respectively). The nonthermal and thermal
components of the warm-target model are presented as solid and dashed curves,
respectively. \label{fig: 20220328_fit}}
\end{figure*}

\begin{figure*}[!ht]
\plotone{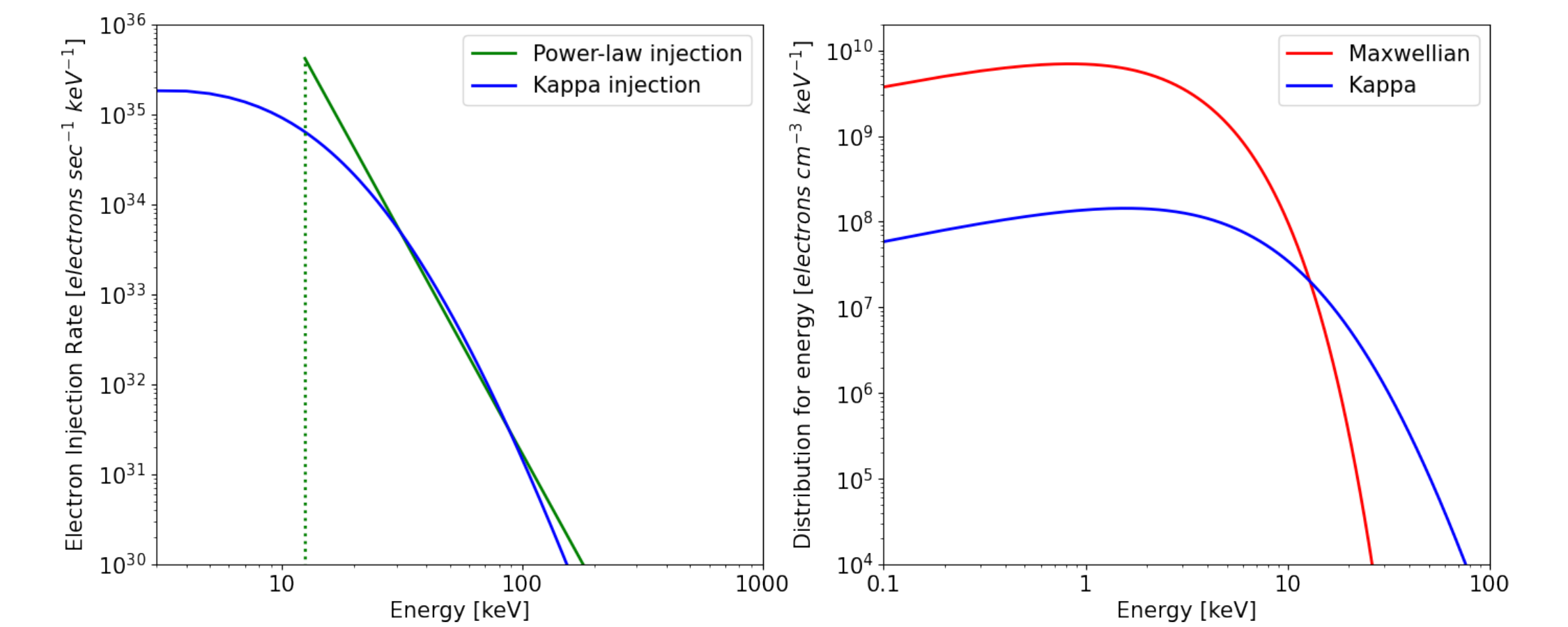}
\caption{Left panel: the injected electron rate spectrum $\dot{N}(E)$ inferred
from warm-target fit results (see Table \ref{tab: 20220328 fit} for fitted
parameters). The green and blue curves correspond to the power-law ($E_c=12.46
\,\rm{keV}$) and kappa forms of electron injection spectrum, respectively. Right
panel: the distribution of electrons for the energy $f_k(E)$ of the fitted kappa
distribution (blue curve, total electron number density $n_k=1.1\times10^{9} \,
\rm{cm}^{-3}$, from Equation \ref{equ: kappa-ele-spec}) and the ambient thermal
plasma Maxwellian distribution (red curve, total electron number density
$n_k=n_\text{th}=2.4\times10^{10} \,\rm{cm}^{-3}$). \label{fig: 20220328_inj}}
\end{figure*}

We conducted the same spectral analysis on the other well-observed GOES-class
M4.0 flare on 2022 March 28. GOES soft X-ray flux peaks at $\sim$ 11:29 UT; the
flare-associated HXR bursts were detected by STIX on board the Solar Orbiter
\citep[Sol-O; ][]{2020A&A...642A...1M}. The STIX temporal profile is shown in
Figure \ref{fig: stix-img}, left panel; with the time difference of 335.7 seconds
between Sol-O and the Earth has been applied to the STIX data. The flare
site is situated in the eastern limb from Sol-O's perspective. The Extreme
Ultraviolet Imager \citep[EUI; ][]{2020A&A...642A...8R}/Full Sun Imager (FSI)
provides EUV context from the same viewpoint (Figure \ref{fig: stix-img}, right
panel). This flare comprises several HXR bursts, and our focus is on the burst
occurring at approximately 11:21 UT (red shaded time range in Figure \ref{fig:
stix-img}, left panel).

The selected burst and preburst time ranges for this flare are
11:21:24--11:21:44 UT and 11:21:12--11:21:17 UT (Figure \ref{fig: stix-img} left
panel, red and black shaded regions), respectively. For this particular event,
bottom pixels have noticeable larger counts than the top pixels. Thus,
the spectrum for fitting and spectroscopic imaging relies solely on the data from
the bottom pixels. The fit range is set to be 6 -- 50~keV. The preburst spectra
yield values of temperature and emission measure of $k_\text{B}T_0=1.67 \,\rm{keV}$
and $EM_0=0.0532\times10^{49} \,\rm{cm}^{-3}$. The right panel of Figure \ref{fig: stix-img} displays X-ray spectroscopic images using the CLEAN algorithm
\citep{1974A&AS...15..417H}, which are overlaid on EUI/FUI 174 \AA\ flare context.
Two loop-top sources (red contours, 8-12 keV), of which the southern one existed
before the flare \citep[See Figure 2 in ][for detailed flare
evolution]{2023A&A...679A..99P}, and two footpoint sources (blue contours, 20-50
keV) are observed. According to the 50\% contour of the two loop-top sources, we
obtain the source size in different dimensions ($r_1=l_1/2=3.4 \,\rm{Mm}$,
$r_2=l_2/2=3.2 \,\rm{Mm}$, pink lines, and $d_1=12.1 \,\rm{Mm}$, $d_1=9.6
\,\rm{Mm}$, red lines). The cross-area (using the northern source, $A=\pi
r_1^2=3.64\times10^{10} \,\rm{cm}^{2}$) and source volume ($V=\pi r_1^2 d_1+\pi
r_2^2 d_2=7.59\times10^{26} \,\rm{cm}^{3}$) can be calculated accordingly. The
constructed semi-elliptical coronal loop (cyan dashed line in Figure \ref{fig:
stix-img} right panel), which is rooted at two HXR footpoint sources and passes
through the loop-top sources, features a half-loop length of $L=18.4 \,\rm{Mm}$.
As a result, we can determine the target loop's thermal properties ($k_\text{B}
T_\text{loop}=k_\text{B}T_0=1.67 \,\rm{keV}$ and
$n_\text{loop}=\sqrt{EM_0/V}=2.4\times10^{10} \, \rm{cm}^{-3}$).

The best-fit results using \textit{f\_vth} and
\textit{f\_thick\_warm\_kappa}/\textit{f\_thick\_warm} are shown in Figure
\ref{fig: 20220328_fit} and Table \ref{tab: 20220328 fit}. To account for STIX
calibration uncertainty, which is still unknown, 6\% systematic error was added
in \texttt{OSPEX} following  
\citet{2024ApJ...964..145J}. From the left panel of Figure \ref{fig: 20220328_inj},
a distinct difference between the kappa and power-law forms of the injection electron
spectrum was noted below 30 keV, consistent with the findings from the 2011
February 24 flare. For this event, the low-energy cutoff $E_c$ obtained from
\textit{f\_thick\_warm} is 12.46~keV. For comparison, the injection rate
$\dot{N}(E=E_c)$ is $0.641$ vs. $4.21$ $\times10^{35} \,\rm{electrons}
\,\rm{sec}^{-1} \,\rm{keV}^{-1}$ for kappa and power-law forms of injection,
respectively. The kappa-form electron distribution, like that for the 2011 February 24
flare, will decrease more rapidly at higher energy ($\kappa>\delta+1$). The
nonthermal power generated by the injected electron spectrum ($3.66$ vs $2.95
\,\times10^{28} \,\rm{erg/s}$, for electrons in power-law and kappa forms of
distribution, respectively) also exhibits significant dissimilarities. The kappa-form electrons generate less nonthermal power ${P_\kappa}/{P_\text{power-law}}
\approx 0.81$ despite larger total injection rates: $20.5$ vs. $13.6$  [$10^{35}
\,\rm{electrons} \,\rm{sec}^{-1}$]. Figure \ref{fig: 20220328_inj} also
illustrates a comparison between the distribution of accelerated kappa-form
electrons (blue curve, total electron number density $n_k=1.1\times10^{9}
\,\rm{cm}^{-3}$) and the ambient thermal Maxwellian distribution (red curve, total
electron number density $n_{th}=2.4\times10^{10} \,\rm{cm}^{-3}$,
$k_\text{B}T=1.67$~keV) in the right panel.

\begin{table}
    \centering
    \begin{tabular}{||c|c|c|c||}
        Parameters & vth+WTM(power-law)  & vth+WTM(kappa) &\\
        \hline
        EM [$10^{49} \,\rm{cm}^{-3}$] & 0.0532 & 0.0532 & Fixed\\
        $k_\text{B} T$ [keV] & 1.67 & 1.67 & Fixed\\
        $n_\text{loop}$ [$10^{10} \,\rm{cm}^{-3}$] & 2.4 & 2.4 & Fixed\\
        $k_\text{B} T_\text{loop}$ [keV] & 1.67 & 1.67 & Fixed\\
         $L$ [Mm] & 18.4 & 18.4 & Fixed\\
        $\dot{N}_0$ [$10^{35} \,\rm{electrons} \,\rm{sec}^{-1}$] & 13.6 $\pm$
        1.3 & 20.5 $\pm$ 1.4 & Free\\
        $\delta$ & 4.86 $\pm$ 0.05 & N/A & Free\\
        $\kappa$ & N/A & 8.85 $\pm$ 0.41 & Free\\
        $E_{c}$ [keV] & 12.46 $\pm$ 0.45 & N/A & Free\\
        $k_\text{B} T_{\kappa}$ [keV] & N/A & 2.97 $\pm$ 0.17 & Free\\
        $\chi^2$ & 0.91 & 1.06 & N/A\\
        Power [$10^{28} \,\rm{erg/s}$] & 3.66 & 2.95 for total; 1.51 above $E_c$
        & N/A\\
    \end{tabular}
    \caption{The best-fit parameters and results of 2022 March 28 M4.0 flare using the warm-target model. The isothermal component and thermal properties of the warm-target model are fixed, and properties associated with the accelerated electrons are set free. The reduced $\chi^2$ and nonthermal electron power $P$ (for kappa-form electrons, the total power and power above power-law low-energy cutoff $E_c$) are also listed.}
    \label{tab: 20220328 fit}
\end{table}

\section{Analysis of the fit parameters and fit results} \label{sec:obs anal} In
this section, we will further investigate how well the three free kappa
parameters ($\dot{N}_0$, $\kappa$, and $T_\kappa$) can be constrained by the fit
with the warm-target model. We will provide the uncertainty of the fit
parameters using different methods. Additionally, we will compare the total
electron density $n_k$ and average electron energy $E_\text{avg}$ derived from
the fit kappa-form spectrum with the corresponding thermal properties.

\subsection{2011 February 24 RHESSI flare} \label{sec:20110224_anal}

\begin{figure*}[!ht]
\plotone{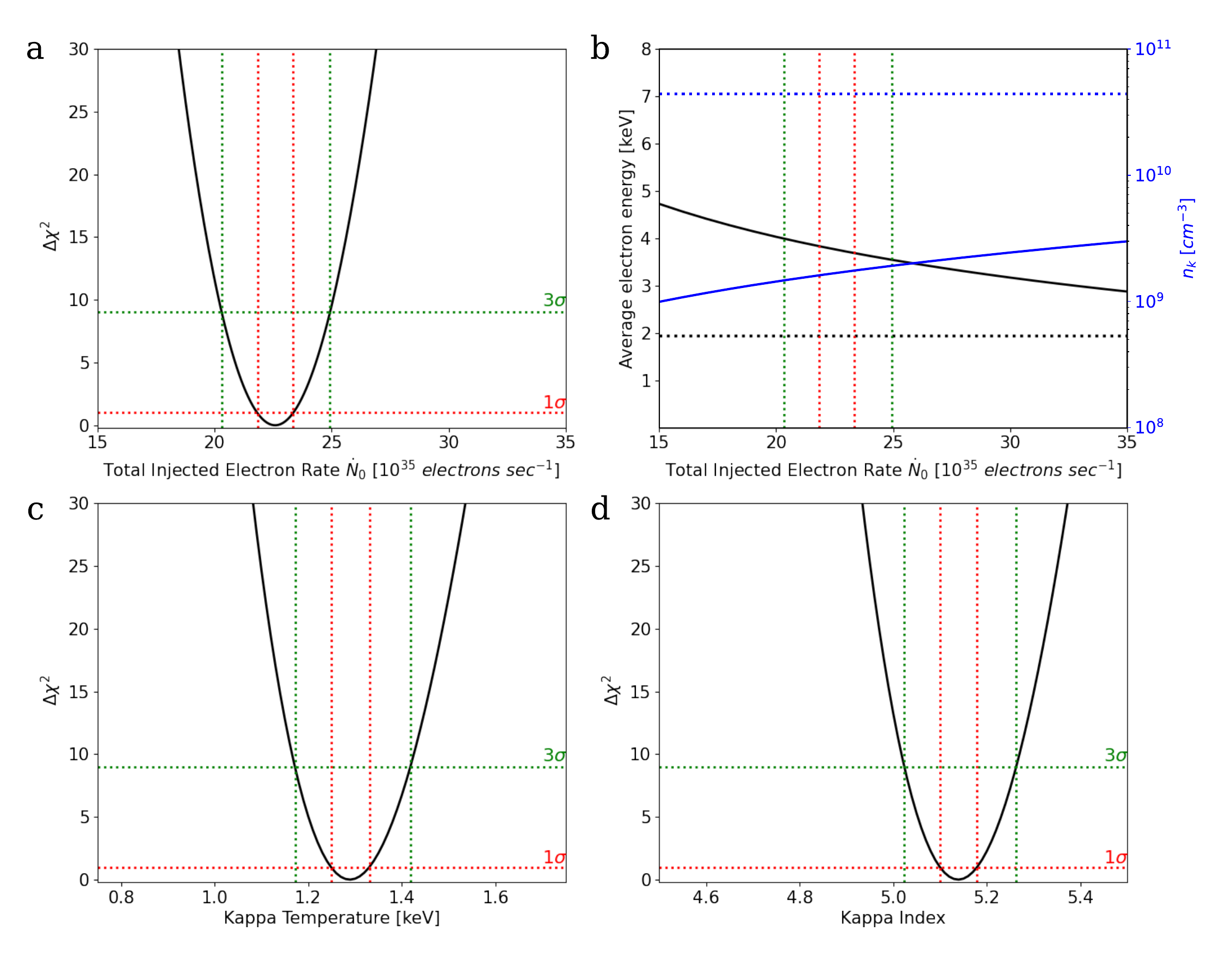}
\caption{(a): $\Delta\chi^2$ against the given total injected electron rate
($\dot{N}_0$) fixed for the fit. The red and green horizontal lines represent
the 1$\sigma$ and 3$\sigma$ uncertainty levels, respectively. Red and green
vertical lines correspond to the $\dot{N}_0$ ranges within the 1$\sigma$ and
3$\sigma$ uncertainty levels, $22.5^{+0.8}_{-0.6}$ $\times10^{35}
\,\rm{electrons} \,\rm{sec}^{-1}$, and $22.5^{+2.4}_{-1.8}$ $\times10^{35}
\,\rm{electrons} \,\rm{sec}^{-1}$, respectively. (b): Corresponding average
electron energy $E_\text{avg}$ (black curve) and number density $n_k$ (blue
curve) derived from the fit kappa-form electron spectrum with different fixed
$\dot{N}_0$ in (a). The red and green vertical lines represent the
$\dot{N}_0$ range within the 1$\sigma$ and 3$\sigma$ uncertainty levels, same
in panel (a). The black horizontal line represents the average electron energy
from the thermal loop ($\frac{3}{2}k_\text{B}T_\text{loop}$) at 1.95 keV. The
blue horizontal line represents the total number density of $4.4\times10^{10}
\,\rm{cm}^{-3}$, which is the number density of ambient coronal loop plasma and
serves as an upper limit of the fit number density. We note that the fit results
within the acceptable range satisfy the two constraints from average electron
energy ($E_{avg-\kappa}>E_{avg-loop}$) and number density ($n_k<n_\text{loop}$).
(c): Same as panel (a) but the fixed parameter is kappa temperature $k_\text{B}T_\kappa$. The
$k_\text{B}T_\kappa$ ranges within the 1$\sigma$ and 3$\sigma$ uncertainty
levels are $1.29^{+0.04}_{-0.04}$~keV and $1.29^{+0.13}_{-0.12}$~keV,
respectively. (d): Same as panel (a) but the fixed parameter is kappa index $\kappa$.
The $\kappa$ ranges within the 1$\sigma$ and 3$\sigma$ uncertainty levels are
$5.14^{+0.04}_{-0.04}$ and $5.14^{+0.12}_{-0.12}$, respectively. \label{fig:
20110224_group}}
\end{figure*}

\begin{figure*}[!ht]
\includegraphics[width=0.8\textwidth]{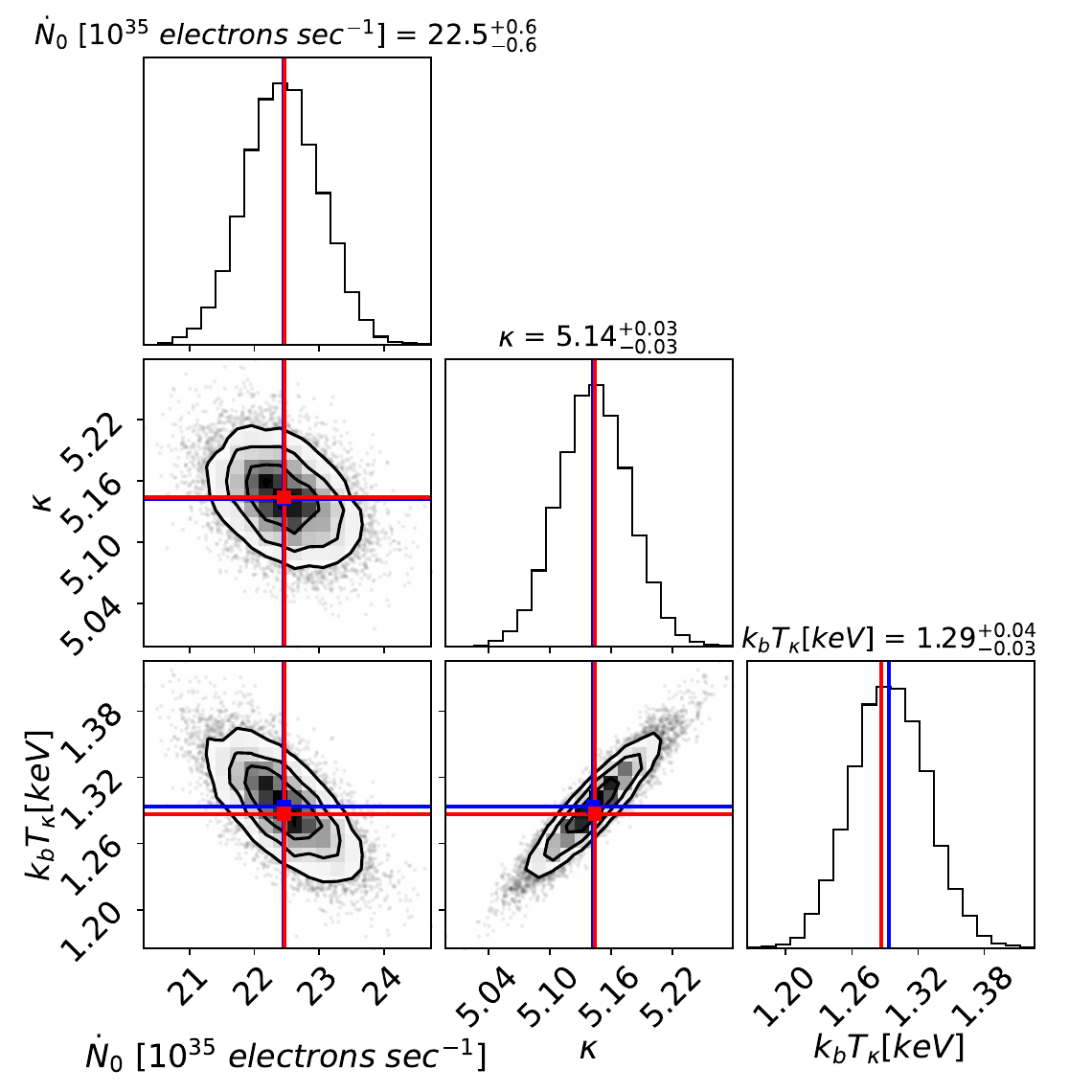}
\centering
\caption{Corner plot of the posterior probability function for the 2011
February 24 M3.5 flare obtained from the Monte Carlo analysis performed using
\texttt{OSPEX}. The diagonal panels display the one-dimensional projected
probability function for each parameter ($\dot{N}_0$, $\kappa$, and $T_\kappa$),
and the remaining panels show the two-dimensional projection. The final fit
results of the Monte Carlo analysis and the best-fit results of the forward fit
are depicted in red and blue, respectively. Additionally, the plot includes the
1$\sigma$ level uncertainty range for each parameter:
$\dot{N}_0$=$22.5^{+0.6}_{-0.6}$ $\times10^{35} \,\rm{electrons}
\,\rm{sec}^{-1}$, $\kappa$=$5.14^{+0.03}_{-0.03}$, and
$k_\text{B}T_\kappa$=$1.29^{+0.04}_{-0.03}$~keV.}
\label{fig: 20110224_mc}
\end{figure*}

\begin{figure*}[!ht]
\centering
\includegraphics[width=0.6\textwidth]{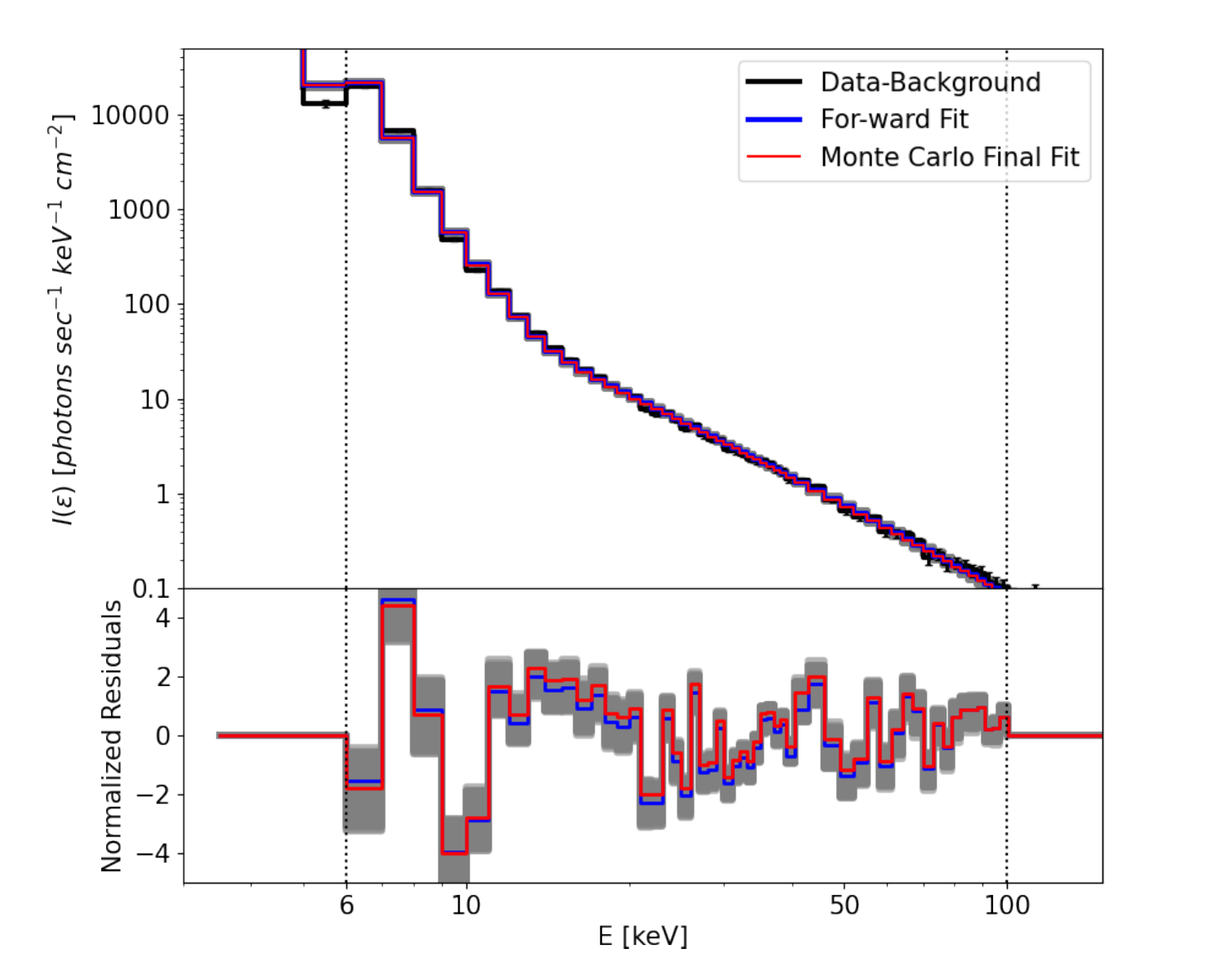}
\caption{The upper panel depicts the photon spectrum of the observation
(background subtracted, black curve), the best-fit results obtained from the
forward fit (blue curve), the final fit result generated by the Monte Carlo
analysis (red curve), and the photon spectrum generated by the Monte Carlo
sampling parameters (gray curves) of the 2011 February 24 M3.5 flare. The lower
panel displays the normalized residuals associated with the corresponding fit
spectrum. \label{fig: 20110224_mc2}}
\end{figure*}

In the previous section, we show that fitting using the functions
\textit{f$\_$vth} and \textit{f\_thick\_warm\_kappa} involves three free
parameters: $\dot{N}_0$, $\kappa$, and $T_\kappa$, all associated with the kappa
distribution. To estimate the uncertainty of each parameter at different
confidence levels, we apply the method used in \citet{2019ApJ...871..225K}. We
began by assessing the uncertainty of total injection rate $\dot{N}_0$. We carry
out spectral fits with $\dot{N}_0$ set at a fixed value while setting the two other
kappa parameters ($\kappa$ and $T_\kappa$) free. By doing this, we can
obtain the best-fit full $\chi^2$ as well as the corresponding kappa index
$\kappa$ and kappa temperature $T_\kappa$ at this fixed $\dot{N}_0$.
Subsequently, we repeat these fits multiple times with different fixed
$\dot{N}_0$ values and record the corresponding full $\chi^2$ values. Further,
we define $\Delta\chi^2=\chi^2-min(\chi^2)$, where $min(\chi^2)$ is the minimum
of all the obtained full $\chi^2$ values. $\Delta\chi^2$ reaches 0 at the
$\dot{N}_0$ value of $22.5$ $\times10^{35} \,\rm{electrons} \,\rm{sec}^{-1}$
(Figure \ref{fig: 20110224_group} (a)), consistent with the best-fit results
shown in Table \ref{tab: 20110224 fit}. The $\Delta\chi^2$--$\dot{N}_0$ curve
(Figure \ref{fig: 20110224_group} (a)) displays a prominent minimum, and the
$\Delta\chi^2$ value changes rapidly in the vicinity of this minimum. The
$\dot{N}_0$ within the 1$\sigma$ uncertainty level (corresponding to a 68\%
confidence level, red horizontal dashed line in Figure \ref{fig: 20110224_group}
(a)) is $22.5^{+0.8}_{-0.6}$ $\times10^{35} \,\rm{electrons}
\,\rm{sec}^{-1}$ (red vertical dashed lines in Figure \ref{fig: 20110224_group}
(a)), and $\dot{N}_0$ within the 3$\sigma$ uncertainty level
(corresponding to a 99\% confidence level, green horizontal dashed line in
Figure \ref{fig: 20110224_group} (a)) is $22.5^{+2.4}_{-1.8}$
$\times10^{35} \,\rm{electrons} \,\rm{sec}^{-1}$ (green vertical dashed lines in
Figure \ref{fig: 20110224_group} (a)). The $\Delta\chi^2$--$\dot{N}_0$ curve
and the estimated uncertainty suggest that $\dot{N}_0$ can be effectively
constrained by the fit with the warm-target model.

The same approach is used to analyze the two other kappa parameters, $\kappa$ and
$T_\kappa$, for uncertainty. Both the $\Delta\chi^2$--$T_\kappa$ (Figure
\ref{fig: 20110224_group} (c)) and $\Delta\chi^2$--$\kappa$ (Figure
\ref{fig: 20110224_group} (d)) curves reveal a distinct minimum, similar to
the $\Delta\chi^2$--$\dot{N}_0$ curve. The ranges of $k_\text{B}T_\kappa$ for
1$\sigma$ and 3$\sigma$ uncertainty levels (red and green horizontal dashed
lines in Figure \ref{fig: 20110224_group} (c)) are $1.29^{+0.04}_{-0.04}$~keV
and $1.29^{+0.13}_{-0.12}$~keV (red and green vertical dashed lines in Figure
\ref{fig: 20110224_group} (c)), respectively. The ranges of $\kappa$ for
1$\sigma$ and 3$\sigma$ uncertainty levels (red and green horizontal dashed
lines in Figure \ref{fig: 20110224_group} (d)) are $5.14^{+0.04}_{-0.04}$ and
$5.14^{+0.12}_{-0.12}$ (red and green vertical dashed lines in Figure \ref{fig:
20110224_group} (d)), respectively.  The small fit uncertainties for all
three kappa parameters indicate that all parameters can be accurately determined
using the warm-target model.

In this study, we utilize the 3$\sigma$ uncertainty level to define the
acceptable range of the fit parameters and investigate the physical properties
within this acceptable range. One valuable property we examine is the average
energy of the electrons $E_\text{avg}$. Here, we first adopt the acceptable
range of $\dot{N}_0$, which falls between $20.3$ and $24.9$ $\times10^{35}
\,\rm{electrons} \,\rm{sec}^{-1}$. Each given $\dot{N}_0$ within the acceptable
range corresponds to a fit kappa-form electron spectrum. The average electron
energy for the kappa-form electron spectrum $E_\text{avg}$ is given by
\begin{equation}
E_\text{avg}  =\frac{\int_{0}^{\infty} Ef_k(E) \,dE}{\int_{0}^{\infty} f_k(E) \,dE} 
=\frac{3}{2}\frac{\kappa}{\kappa-2.5}k_\text{B} T_\kappa. 
\label{equ: E_avg kappa}
\end{equation}
For comparison, the average electron energy of the Maxwellian electron spectrum
is $E_\text{avg}=\frac{3}{2}k_\text{B} T$. The obtained
$E_\text{avg}$--$\dot{N}_0$ curve is depicted by the black curve in Figure
\ref{fig: 20110224_group} (b). It is worth noting that, within the
acceptable range (green vertical dashed lines in Figure \ref{fig:
20110224_group} (b)), the average electron energies derived from the
kappa-form electron spectrum $E_{avg-\kappa}$ are consistently higher than the
average electron energy of the thermal plasma in the target loop
($E_{avg-loop}=\frac{3}{2}k_\text{B} T_\text{loop}=$1.95~keV, black horizontal
dashed lines in Figure \ref{fig: 20110224_group} (b)). According to Equation
\ref{equ: kappa-ele-spec}, the total electron number density $n_k$ can be
determined once the injection area $A$ is known. We use the cross-area of the
loop-top X-ray coronal source ($A=\pi r^2=4.21\times10^{10} \, \rm{cm}^{2}$,
red outlier contour in Figure \ref{fig: rhe-img} right panel) as the injection
area $A$ to calculate the total electron number density $n_k$. The
$n_k$--$\dot{N}_0$ curve is shown as the blue curve in Figure \ref{fig:
20110224_group} (b). Within the acceptable range of $\dot{N}_0$, the total
electron number density $n_k$ is found to be lower than the thermal electron
density of the target coronal loop $n_\text{loop}$ (blue horizontal dashed lines
in Figure \ref{fig: 20110224_group} (b), $4.4\times10^{10} \,\rm{cm}^{-3}$). While the electron injection area cannot be precisely determined, and it may be
smaller than the observed loop cross-area. $n_k$ is unlikely to be higher than
the loop density $n_\text{loop}$, which requires an extremely small injection
area $A<1.60\times10^{9} \,\rm{cm}^{2}$. The results indicate that the values
of $E_{avg-\kappa}$ and $n_k$ are physically reasonable when compared with the
ambient thermal plasma. The fitted kappa-form electron spectrum and associated
properties ($E_{avg-\kappa}$ and $n_k$) within the acceptable range for
$T_\kappa$ and $\kappa$ closely resemble those obtained within the acceptable
range for $\dot{N}_0$, and are consistent with the previous findings. Consequently,
we have opted not to display them in the plot.

In order to further evaluate how well the fit parameters can be determined and
cross-check their uncertainty, we utilized the Monte Carlo analysis provided by
\texttt{OSPEX}. Detailed information on this method is available in the 'Fit
Parameter Uncertainty Analysis' section of \citet{2020ascl.soft07018T}. In
addition, \citet{2013ApJ...769...89I} discussed commonly used methods, including
the Monte Carlo method, for estimating the constraint of fit parameters. The
Monte Carlo analysis requires a predetermined best-fit result, which is
available in Table \ref{tab: 20110224 fit}. The Monte Carlo sampling explores
the multidimensional parameter space ($\dot{N}_0$, $\kappa$, and $T_\kappa$)
surrounding the best-fit results by assuming a Gaussian probability function for
each parameter. Figure \ref{fig: 20110224_mc} shows the posterior probability
density function for the three free parameters obtained from the Monte Carlo
analysis. The final fit results of the Monte Carlo analysis, as well as the
predetermined best-fit results from the forward fit, are also depicted by red
and blue dots, respectively. The diagonal panels of the corner plot show the
one-dimensional projection of the probability density function. For the 2011
February 24 M3.5 flare, each parameter exhibits a prominent peak and a
relatively narrow width, indicating that they are well constrained around the
best-fit results. The 1$\sigma$ uncertainty range for each parameter is
shown in Figure \ref{fig: 20110224_mc}: $\dot{N}_0$=$22.5^{+0.6}_{-0.6}$
$\times10^{35} \,\rm{electrons} \,\rm{sec}^{-1}$,
$\kappa$=$5.14^{+0.03}_{-0.03}$, and
$k_\text{B}T_\kappa$=$1.29^{+0.04}_{-0.03}$~keV, comparable with the uncertainty
obtained from the previous method. The nondiagonal panels display the
probability density function projected for different parameter pairs, and
illustrate the correlations between different parameters. We observed that
$\kappa$ and $T_\kappa$ are positively correlated, while $\dot{N}_0$ is
anticorrelated with $\kappa$ and $T_\kappa$. The photon spectrum and normalized
residuals generated by Monte Carlo sampling, the Monte Carlo final fit result, and the best forward fit result are shown as gray, red, and blue curves in Figure
\ref{fig: 20110224_mc2}.

\subsection{2022 March 28 STIX flare} \label{sec:20220328_anal}

\begin{figure*}[!ht]
\plotone{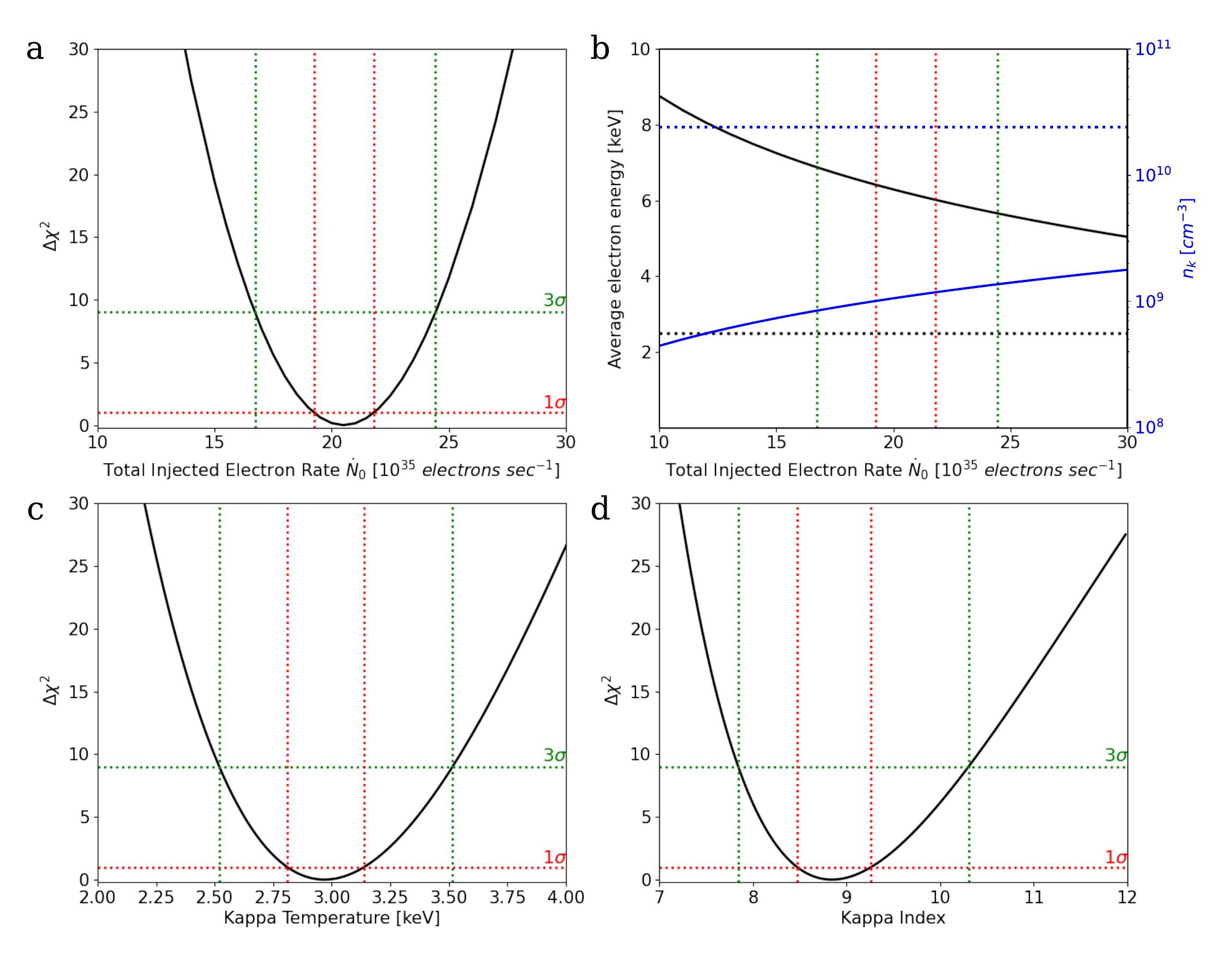}
\caption{Same as Figure \ref{fig: 20110224_group} but for the 2022 March 28 M4.0
flare. (a): $\Delta\chi^2$ against the different total injected electron
rate ($\dot{N}_0$). The red and green horizontal lines represent the 1$\sigma$
and 3$\sigma$ uncertainty levels, respectively. Red and green vertical lines
correspond to the $\dot{N}_0$ range within the 1$\sigma$ and 3$\sigma$
uncertainty levels, $20.5^{+1.3}_{-1.2}$ $\times10^{35} \,\rm{electrons}
\,\rm{sec}^{-1}$ and $20.5^{+3.9}_{-3.8}$ $\times10^{35} \,\rm{electrons}
\,\rm{sec}^{-1}$, respectively. (b): Corresponding average electron energy
$E_\text{avg}$ (black curve) and number density $n_k$ (blue curve) derived from
the fit kappa-form electron spectrum for different fixed $\dot{N}_0$. The red and
green vertical lines represent the $\dot{N}_0$ range within the 1$\sigma$ and
3$\sigma$ uncertainty levels, the same as in panel (a). The black horizontal line
represents the average electron energy from the thermal loop
($\frac{3}{2}k_\text{B}T_\text{loop}$) at 2.51 keV. The blue horizontal line
represents the electron number density of the ambient thermal plasma,
$2.4\times10^{10} \,\rm{cm}^{-3}$. (c): Same as panel (a) but the fixed
parameter is kappa temperature. The $k_\text{B}T_\kappa$ ranges within the
1$\sigma$ and 3$\sigma$ uncertainty levels are $2.97^{+0.17}_{-0.16}$~keV and
$2.97^{+0.55}_{-0.45}$~keV, respectively. (d): Same as panel a but the fixed
parameter is kappa index. The $\kappa$ ranges within the 1$\sigma$ and
3$\sigma$ uncertainty levels are $8.85^{+0.41}_{-0.38}$ and
$8.85^{+1.45}_{-1.01}$, respectively. \label{fig: 20220328_group}}
\end{figure*}

\begin{figure*}[!ht]
\centering
\includegraphics[width=0.8\textwidth]{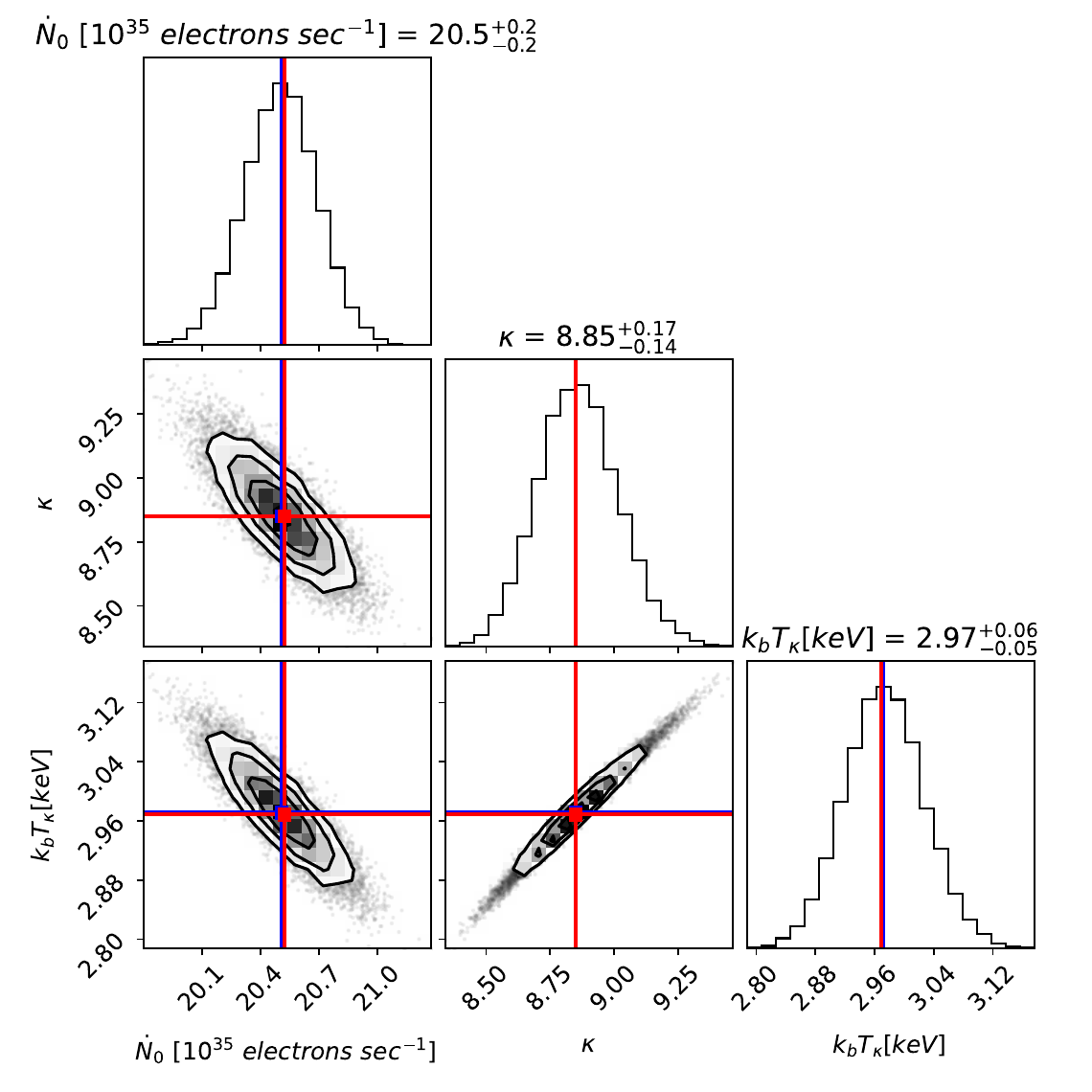}
\caption{Corner plot of the posterior probability function for the 2022
March 28 M4.0 flare obtained from the Monte Carlo analysis performed using
\texttt{OSPEX}. As in Figure \ref{fig: 20110224_mc}, the diagonal panels
display the one-dimensional projected probability function for each parameter
($\dot{N}_0$, $\kappa$, and $T_\kappa$), and the nondiagonal panels show the
two-dimensional projection. The final fit results of the Monte Carlo analysis
and the best-fit results of the forward fit are depicted in red and blue,
respectively. The plot also includes the 1$\sigma$ level uncertainty range for
each parameter: $\dot{N}_0$=$20.5^{+0.2}_{-0.2}$ $\times10^{35} \,\rm{electrons}
\,\rm{sec}^{-1}$, $\kappa$=$8.85^{+0.17}_{-0.14}$, and
$k_\text{B}T_\kappa$=$2.97^{+0.06}_{-0.05}$~keV. \label{fig: 20220328_mc}}
\end{figure*}

\begin{figure*}[!ht]
\centering
\includegraphics[width=0.6\textwidth]{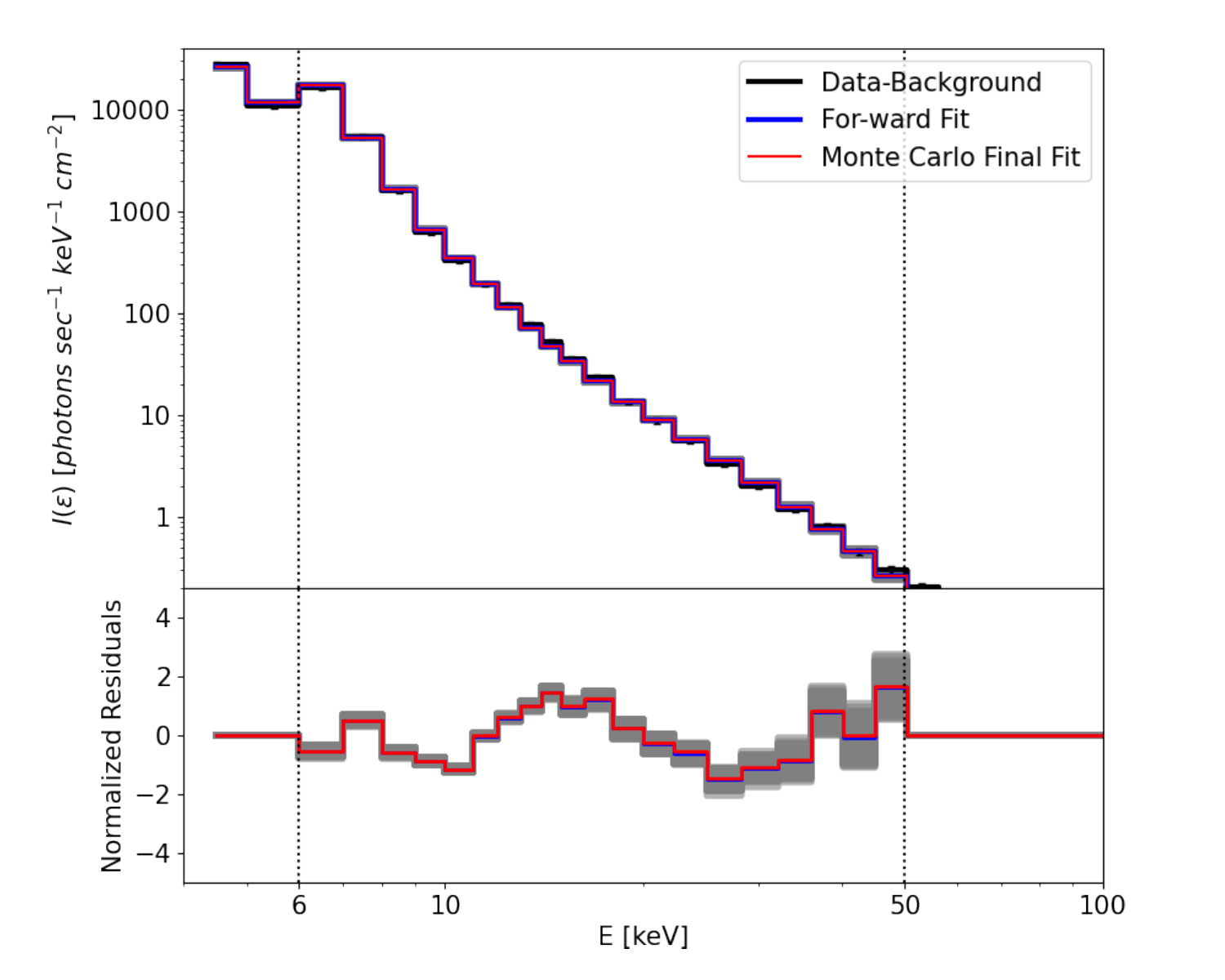}
\caption{Same as Figure \ref{fig: 20110224_mc2}, but for the 2022 March 28 M4.0
flare. The upper panel shows the photon spectrum of the observation (background
subtracted, black curve), the best-fit results obtained from the forward fit
(blue curve), the final fit result generated by the Monte Carlo analysis (red
curve), and the photon spectrum generated by the Monte Carlo sampling parameters
(gray curves). The lower panel displays the normalized residuals associated with
the corresponding fit spectrum.  
\label{fig: 20220328_mc2}}
\end{figure*}

We conducted the same analysis on the 2022 March 28 M4.0 flare observed by STIX.
Similar to the 2011 February 24 M3.5 flare, the $\Delta\chi^2$--$\dot{N}_0$,
$\Delta\chi^2$--$T_\kappa$, and $\Delta\chi^2$--$\kappa$ curves (black curves in
Figures \ref{fig: 20220328_group} (a), (c), and (d), respectively) all exhibit a
prominent minimum. The ranges for each parameter at 1$\sigma$ and 3$\sigma$
uncertainty levels are as follows: for $\dot{N}_0$, $20.5^{+1.3}_{-1.2}$
$\times10^{35} \,\rm{electrons} \,\rm{sec}^{-1}$, and $20.5^{+3.9}_{-3.8}$
$\times10^{35} \,\rm{electrons} \,\rm{sec}^{-1}$; for $k_\text{B}T_\kappa$,
$2.97^{+0.17}_{-0.16}$~keV, and $2.97^{+0.55}_{-0.45}$~keV; for $\kappa$,
$8.85^{+0.41}_{-0.38}$, and $8.85^{+1.45}_{-1.01}$. The spectral fit for the
2022 March 28 M4.0 flare, due to a narrow fit range (6--50~keV) and a 6\%
systematic error from STIX (2\% for RHESSI), results in relatively larger
uncertainties, particularly for $T_\kappa$ and $\kappa$. However, these
uncertainties still demonstrate that the warm-target fit effectively constrains
the fit parameters. We adopt the 3$\sigma$ uncertainty level as the acceptable
range, consistent with our approach for the 2011 February 24 M3.5 flare. The
derived average electron energy $E_\text{avg}$ and total electron density $n_k$
both satisfy $E_{avg-\kappa}>E_{avg-loop}$ and $n_k<n_\text{loop}$ within the
acceptable range (Figure \ref{fig: 20220328_group} (b)). Additionally, the results of
Monte Carlo analysis are displayed in Figures \ref{fig: 20220328_mc} and
\ref{fig: 20220328_mc2}. The 1$\sigma$ uncertainty level ranges for each
parameter are: $\dot{N}_0$=$20.5^{+0.2}_{-0.2}$ $\times10^{35} \,\rm{electrons}
\,\rm{sec}^{-1}$, $\kappa$=$8.85^{+0.17}_{-0.14}$, and
$k_\text{B}T_\kappa$=$2.97^{+0.06}_{-0.05}$~keV. The correlations between these
parameters are consistent with the findings from the 2011 February 24 M3.5 flare
(Figure \ref{fig: 20220328_mc} diagonal panels).

\section{Summary} \label{sec:discussion} 

In this study, we utilized the warm-target model to analyze two GOES M-class limb
flares. The M3.5 class flare occurred on 2011 February 24 and was observed by
RHESSI, while STIX observed the other M4.0 class flare on 2022 March 28. We
used the warm-target model due to its ability to accurately determine nonthermal
electron properties, such as nonthermal power, which can hardly be constrained
by the commonly used cold-target model. Unlike previous investigations, 
the kappa distribution is used to
characterize the accelerated/injected electrons. 
Unlike the power-law distribution, the kappa distribution is finite
as the electron speed approaches zero, 
thereby effectively covering the entire electron energy range without the
requirement of an arbitrary low-energy cutoff. 
Consequently, the fitted
kappa-form electron spectrum can characterize electrons whose kinetic
energy falls below the X-ray instrument's sensitive range. 
The results of the fitting and subsequent analysis are summarized below.

\begin{itemize}

\item The best-fit results using the warm-target model in kappa-form injected
electrons feature reasonable electron spectrum. For comparison, we also employed
the warm-target model in the power-law electron distribution to fit the HXR
spectrum. We found that the obtained power-law-form and kappa-form electron spectra
generate similar photon spectra within the fitting range. However, the
kappa-form electron spectrum produces less nonthermal power despite having a
comparable or higher total injection rate $\dot{N}_0$. This energy difference
is due to the behavior of the electron spectrum. The obtained
kappa-form and power-law-form electron spectra exhibited similar behavior at the
deka-keV level, while the kappa-form spectrum displayed a significantly lower
flux below $\sim$30~keV. Moreover, the kappa-form electron spectrum decreased
more rapidly than the power-law spectrum ($\kappa>\delta+1$) at higher energies
(typically above 100~keV, out of the fit range).

\item The fit with the warm-target model in kappa-form electron injection
involves three free parameters ($\dot{N}_0$, $\kappa$, and $T_\kappa$). 
The $\Delta\chi^2$-$\dot{N}_0$ curve reveals a prominent minimum, 
suggesting an accurate determination of $\dot{N}_0$ within a narrow range. 
For the M3.5 flare on 2011 February 24, the range of $\dot{N}_0$ within the 1$\sigma$ 
uncertainty level is 
$22.5^{+0.8}_{-0.6}$ $\times10^{35} \,\rm{electrons} \,\rm{sec}^{-1}$, and within the
3$\sigma$ uncertainty level range it is $22.5^{+2.4}_{-1.8}$ $\times10^{35}
\,\rm{electrons} \,\rm{sec}^{-1}$. For the M4.0 flare on 2022 March 28,
the range of $\dot{N}_0$ within the 1$\sigma$ uncertainty level is $20.5^{+1.3}_{-1.2}$
$\times10^{35} \,\rm{electrons} \,\rm{sec}^{-1}$, and within the 3$\sigma$
uncertainty level it is $20.5^{+3.9}_{-3.8}$ $\times10^{35} \,\rm{electrons}
\,\rm{sec}^{-1}$. We apply the same analyses to the other two fit parameters,
$\kappa$ and $T_\kappa$, for their uncertainty. 
For the M3.5 flare on 2011 February 24, $\kappa$
within 1$\sigma$ and 3$\sigma$ uncertainty levels is
$5.14^{+0.04}_{-0.04}$ and $5.14^{+0.12}_{-0.12}$, respectively.
$k_\text{B}T_\kappa$ within 1$\sigma$ and 3$\sigma$ uncertainty levels is
$1.29^{+0.04}_{-0.04}$~keV and $1.29^{+0.13}_{-0.12}$~keV, respectively. For the
M4.0 flare on March 28, 2022, $\kappa$ within 1-$\sigma$ and 3-$\sigma$
uncertainty level range is $8.85^{+0.41}_{-0.38}$ and $8.85^{+1.45}_{-1.01}$,
respectively. $k_\text{B}T_\kappa$ within 1-$\sigma$ and 3-$\sigma$ uncertainty
level range is $2.97^{+0.17}_{-0.16}$~keV and $2.97^{+0.55}_{-0.45}$~keV,
respectively. All the three parameters, $\dot{N}_0$, $T_\kappa$ and $\kappa$ are
tightly constrained within a narrow range. Additionally, we carry out a Monte
Carlo analysis to further confirm the effectiveness in determining the fit
parameters with the warm-target model. The resulting posterior probability
density function for all three kappa parameters demonstrates distinct peaks and
relatively narrow widths, indicating well-constrained fit parameters. The Monte
Carlo analysis also provides a 1$\sigma$ level uncertainty for each parameter:
for the 2011 February 24 M3.5 flare, $\dot{N}_0$=$22.5^{+0.6}_{-0.6}$ $\times10^{35}
\,\rm{electrons} \,\rm{sec}^{-1}$, $\kappa$=$5.14^{+0.03}_{-0.03}$, and
$k_\text{B}T_\kappa$=$1.29^{+0.04}_{-0.03}$~keV; for the 2022 March 28 M4.0 flare,
$\dot{N}_0$=$20.5^{+0.2}_{-0.2}$ $\times10^{35} \,\rm{electrons}
\,\rm{sec}^{-1}$, $\kappa$=$8.85^{+0.17}_{-0.14}$, and
$k_\text{B}T_\kappa$=$2.97^{+0.06}_{-0.05}$~keV.

\item Unlike the power-law distribution, which requires a low-energy cutoff
$E_c$, the kappa-form electron spectrum works for all energies, anchored at zero
speed and extending to the speed of light. Therefore, the kappa distribution
facilitates the determination of the accelerated electron number density $n_k$
(with the information of the injection area $A$) and the average electron energy
$E_\text{avg}$. For both flares in this study, we find that the derived total electron energy density is lower
than the thermal number density of the target loop ($n_k<n_\text{loop}$,
$1.7\times10^{9} \,\rm{cm}^{-3}<4.4\times10^{10} \,\rm{cm}^{-3}$ and
$1.1\times10^{9} \,\rm{cm}^{-3}<2.4\times10^{10} \,\rm{cm}^{-3}$ for 2011
February 24 and 2022 March 28 flares, respectively), while the average electron
energy surpasses that of the ambient Maxwellian thermal plasma (
$E_{avg-\kappa}>E_{avg-loop}$, 3.77 keV $>$ 1.95 keV and 6.21 keV $>$ 2.51 keV
for 2011 February 24 and 2022 March 28 flares, respectively).

\end{itemize}

In this study, we have demonstrated that the fit with the warm-target model
provides a convincing kappa-form accelerated electron spectrum. The associated
kappa distribution parameters are valuable for understanding electron
acceleration and transport. Here, we follow the model proposed by
\citet{2014ApJ...796..142B}, which uses the Fokker-Planck equation to analyze
the evolution of flare-associated electrons. \citealt{2014ApJ...796..142B}
proposed that the kappa distribution arises from stochastic acceleration in the
presence of Coulomb collisions and velocity diffusion. The kappa index $\kappa$
is given by $\kappa=\frac{\tau_{acc}(v)}{2\tau_{c}(v)}$. The collisional
deceleration time can be estimated as: $\tau_{c}(v)\approx{v^3}/{\Gamma}$, where
the collisional parameter $\Gamma={2Kn}/{m_e^2}$. Here, we take the parameters
from the 2011 February 24 M3.5 flare event for study. For the upper limit of
electron energy $E_l$ enabling the complete stop within the target, we have
$E_l=\sqrt{2KnL}\approx19.0\, \rm{keV}$. At $E_l$, we have the electron speed
$v_l=8.0\times10^{9}$ $\rm{cm/s}$ and the collisional deceleration time
$\tau_{c}(v_l)\approx0.71$ $\rm{seconds}$. Thus the acceleration time can be
determined as $\tau_\text{acc}(v_l)=2\kappa \tau_{c}(v_l)\approx7.33$
$\rm{seconds}$. The collisional diffusion time
$\tau_{d}(v_l)=\frac{2v_l^5}{\Gamma v_{te}^2}\approx19.77$ $\rm{seconds}$, where
thermal speed $v_\text{te}=\sqrt{2k_\text{B}T_\kappa}/m_e\approx2.1\times10^{9}$
$\rm{cm/s}$.

The ratio between the density of accelerated nonthermal electrons
($n_\text{nth}$) and the density of ambient thermal electrons ($n_\text{th}$) can provide
valuable insights into electron acceleration during solar flares. Previous
studies have shown that the ratio of nonthermal electrons to protons
($n_\text{p}=n_\text{nth}+n_\text{th}$ in the hydrogen plasma) is of the order
of approximately 1\% \citep{2013ApJ...764....6O}. In this study, we determined
the density of accelerated nonthermal electrons in the corona by integrating the
kappa electron spectrum over 10~keV($n_\text{nth}[>10~
\mbox{keV}]\sim1.2\times10^{8} \,\rm{cm}^{-3}$, and $2.0\times10^{8}
\,\rm{cm}^{-3}$ for the 2011 February 24 M3.5 and 2022 March 28 M4.0 flares,
respectively). For the thermal electron density at the flare site, we take the
loop electron density obtained from the X-ray thermal component
($n_\text{th}=n_\text{loop}$). The ratio between the density of coronal
accelerated nonthermal electrons and ambient protons is found to be less than
1\% (${n_\text{nth}}/{n_\text{p}}\sim0.003$ and 0.008, for the 2011 February 24
M3.5 and 2022 March 28 flares, respectively), consistent with recent studies
\citep{2023ApJ...947L..13K}.

Here, we note that the electron number density is calculated using Equation \ref{equ:
kappa-ele-spec} and is dependent on the injection area $A$. We estimated the
injection area $A$ using 50\% of the loop-top X-ray source contour. The
resulting total injected electron density is approximately an order of magnitude
lower than the estimated coronal loop thermal electron density ($1.7\times10^{9}
\,\rm{cm}^{-3}<4.4\times10^{10} \,\rm{cm}^{-3}$ and $1.1\times10^{9}
\,\rm{cm}^{-3}<2.4\times10^{10} \,\rm{cm}^{-3}$ for the 2011 February 24 and 2022
March 28 flares, respectively). 
It is important to note that the electron injection
in the coronal region may be nonuniform in height.
According to \citet{2015A&A...584A..89J}, 
the actual injection might be smaller.
In this study, we are utilizing the spatially integrated spectrum for analysis. At present, the signal-to-noise ratio does not sufficiently support precise spatially resolved spectral diagnostics at the arcsecond scale. Once we obtain the spatially resolved spectral information from the improved X-ray observations, the electron density derived from the kappa distribution will be more persuasive.

Since both RHESSI and Sol-O/STIX are not sensitive to photons below 
$\sim$3-6~keV, the thermal properties obtained from
RHESSI X-ray spectral diagnostics could differ from those obtained from SDO/AIA EUV
diagnostics \citep{2013ApJ...779..107B}. Furthermore, thermal properties 
inferred from the observed X-ray spectrum are likely to be height-dependent \citep{2015A&A...584A..89J}. The obtained kappa-form energetic electrons 
allows for a more comprehensive combined study of the X-ray and EUV diagnostics. 
Potential future applications of the derived kappa distribution also include studying 
acceleration mechanisms based on the average electron energy and nonthermal power. 

\begin{acknowledgments}
The authors thank the referee for improving the manuscript. The work was supported via the STFC/UKRI grant ST/T000422/1.
\end{acknowledgments}

\bibliography{sample631}{}
\bibliographystyle{aasjournal}

\end{document}